\documentclass[10pt,journal,compsoc]{IEEEtran}
\usepackage{booktabs} 
\usepackage{soul}
\usepackage{amssymb}
\usepackage{amsfonts,amsmath}
\usepackage{mathrsfs}
\usepackage[ruled, linesnumbered]{algorithm2e}
\usepackage[noend]{algpseudocode}
\usepackage{amsthm}
\usepackage{bbm}
\usepackage{caption}
\usepackage{multirow}
\usepackage{subcaption}
\usepackage{comment}
\usepackage{graphicx}
\usepackage{color}
\usepackage{enumerate}
\usepackage{caption}
\usepackage{hyperref}
\usepackage{url}
\usepackage{lipsum, babel, multicol}
\usepackage{listings}
\usepackage{amssymb}
\usepackage{pifont}

\makeatletter

\newcommand{\removelatexerror}{\let\@latex@error\@gobble}
\makeatother

\newtheorem{insight}{Insight}

\newtheorem{assumption}{Assumption}

\newcommand{\xmark}{\ding{55}}
\definecolor{dkgreen}{rgb}{0,0.6,0}
\definecolor{mauve}{rgb}{0.58,0,0.82}

\newcommand{\ignore}[1]{}

\DeclareMathOperator{\sign}{sign}
\DeclareMathOperator{\Proj}{Proj}
\DeclareMathOperator*{\argmax}{arg\,max}
\DeclareMathOperator*{\argmin}{arg\,min}
\DeclareMathOperator*{\minimum}{minimum}
\DeclareMathOperator*{\maximum}{maximum}
\DeclareMathOperator*{\softmax}{softmax}

\ifCLASSOPTIONcompsoc
  \usepackage[nocompress]{cite}
\else
  \usepackage{cite}
\fi

\ifCLASSINFOpdf
  
\else
  
\fi

\begin{document}

\title{A Framework for Enhancing Deep Neural Networks Against Adversarial Malware\thanks{A preliminary version of the paper was presented at AICS'2019, which does not publish any formal proceedings \cite{DBLP:journals/corr/abs-1812-08108}.}}

\author{Deqiang Li,
	Qianmu Li,
	Yanfang Ye,
	and Shouhuai Xu
	\IEEEcompsocitemizethanks{\IEEEcompsocthanksitem D. Li is with School of Computer Science and Engineering, Nanjing University of Science and Technology. 
		\IEEEcompsocthanksitem Q. Li is with School of Computer Science and Engineering, Nanjing University of Science and Technology and School of Intelligent Manufacturing, Wuyi University
		\IEEEcompsocthanksitem Y. Ye is with Department of Computer and Data Sciences, Case Western Reserve University. 
		\IEEEcompsocthanksitem S. Xu is with Department of Computer Science, University of Colorado Colorado Springs. This work was done when he was at University of Texas at San Antonio, USA. 
		E-mail: sxu@uccs.edu
	}
}

\markboth{IEEE Transactions on Network Science and Engineering}%
{D. Li, Q. Li, Y. Ye, and S. Xu: A Framework for Enhancing Deep Neural Networks Against Adversarial Malware}

\IEEEtitleabstractindextext{%
\begin{abstract}
Machine learning-based malware detection is known to be vulnerable to adversarial evasion attacks.
The state-of-the-art is that there are no effective defenses against these attacks. As a response to the adversarial malware classification challenge organized by the MIT Lincoln Lab and associated with the AAAI-19 Workshop on Artificial Intelligence for Cyber Security (AICS'2019), we propose six guiding principles to enhance the robustness of deep neural networks. Some of these principles have been scattered in the literature, but the others are introduced in this paper for the first time. Under the guidance of these six principles, we propose a defense framework to enhance the robustness of deep neural networks against adversarial malware evasion attacks. By conducting experiments with the Drebin Android malware dataset, we show that the framework can achieve a 98.49\% accuracy (on average) against grey-box attacks, where the attacker knows some information about the defense and the defender knows some information about the attack, and an 89.14\% accuracy (on average) against the more capable white-box attacks, where the attacker knows everything about the defense and the defender knows some information about the attack. The framework wins the AICS'2019 challenge by achieving a 76.02\% accuracy, where neither the attacker (i.e., the challenge organizer) knows the framework or defense nor we (the defender) know the attacks. This gap highlights the importance of knowing about the attack.
\end{abstract}

}

\IEEEpubid{2327-4697 © 2020 IEEE. Personal use is permitted, but republication/redistribution requires IEEE permission.}
\maketitle

\IEEEdisplaynontitleabstractindextext

\IEEEpeerreviewmaketitle

\IEEEraisesectionheading{\section{Introduction}\label{sec:introduction}}

\IEEEPARstart{M}{alware} remains a big threat to cyber security despite communities' tremendous countermeasure efforts. For example, Symantec \cite{symantec:Online} reports seeing 357,019,453 new malware variants in the year 2016, 669,974,865 in the year 2017, and 246,002,762 in the year 2018.
Worse yet, there is an increasing number of malware variants that attempted to undermine anti-virus tools and indeed evaded many malware detection systems \cite{cisco:Online}.

In order to cope with the increasingly severe situation, we have to resort to machine learning for automating the detection of malware in the wild \cite{DBLP:journals/csur/YeLAI17}. However, machine learning based techniques are vulnerable to adversarial evasion attacks, by which an adaptive attacker perturbs or manipulates malware examples into adversarial examples that would be detected as benign rather than malicious (see, for example, \cite{grosse2017adversarial,DBLP:conf/eisic/ChenYB17,al2018adversarial,DBLP:conf/ijcai/HouYSA18,pierazzi2020intriguing,li2020adversarial, kucuk2020deceiving}).
The state-of-the-art is that there are many attacks, but the problem of effective defense is largely open. For example, adversarial training is known to be able to harden classifiers against adversarial examples, but requires knowing about the attack in terms of (for example) its manipulation set \cite{al2018adversarial}. This is indeed the context in which the AICS'2019 Malware Classification Challenge is proposed. In a broader context, adversarial malware examples are a particular kind of attacks against adversarial machine learning. Although adversarial machine learning has received much attention in application domains such as image processing (see, e.g., \cite{szegedyZSBEGF13,goodfellow6572explaining,serban2018adversarial}), the problem of adversarial malware examples are much less investigated \cite{grosse2017adversarial,al2018adversarial,pierazzi2020intriguing,li2020sok}.

The AICS'2019 challenge mentioned above is essentially about {\em whether we can defend against adversarial examples in the wild}. The challenge is characterized as follows. First, we (i.e., any team participating in the Challenge as the defender) are given a training set in the form of anonymized feature representation by the Challenge organizer (i.e., we do not even know what the feature names are), as well as the corresponding ground-truth labels. We are informed by the Challenge organizer that the training data contains {\em no} adversarial examples.
Second, we are given a set of test data (again, in anonymized feature representation) and are told that the test data contains both adversarial examples and non-adversarial examples. We do not know what attacks are used by the Challenge organizers. We do not know which examples in the test set are perturbed by adversaries. 
This means that we neither know which are adversarial examples, nor the attacks that are used to generate them, nor the manipulation set.
Third, our task is to accurately classify the test data, including both adversarial and non-adversarial examples.
The setting of the Challenge is realistic because in the real world defenders do not know attacker's specifications such as attack methods, manipulation sets, and specific adversarial examples. The importance of the problem in defending against adversarial malware examples and the realistic setting of the Challenge motivate the present study.

\subsection{Our Contributions}

In this paper, we make the following contributions. First, we propose, to the best of our knowledge, the first systematic defense framework to enhance the robustness of Deep Neural Network (DNN)-based malware classifiers against adversarial evasion attacks. The framework is designed under the guidance of a set of principles, some of which are known but scattered in the literature (e.g., using an ensemble of classifiers and {\em minmax} adversarial training), but others are introduced in this paper for the first time, such as the following. We propose (i) using 
white-box attack, where the attacker knows everything about the defense, to bound the capability of grey-box attacks with respect to the $\ell_p~(p \geq 1)$ norm, where the attacker knows something about the defense; (ii) using adversarial regularization (i.e., adversarial training with small perturbations) when the manipulation set is not available to the defender; (iii) leveraging semantics-preserving representations (realized by Denoising Auto-Encoder or DAE for shorthand).

Second, we empirically validate the effectiveness of the framework against 11 grey-box attacks and 9 white-box attacks (i.e., 20 attacks in total). The 11 grey-box attacks 
include the Random attack, two Mimicry attacks \cite{7917369}, the Fast Gradient Sign Method (FGSM) attack \cite{goodfellow6572explaining}, Grosse attack \cite{grosse2017adversarial}, Bit Gradient Ascent (BGA) attack \cite{al2018adversarial}, Bit Gradient Ascent (BCA) attack \cite{al2018adversarial}, and four variants of the Projected Gradient Descent (PGD) attacks. The 9 white-box attacks leverage the victim models directly and the attack algorithms are the same as the latter 9 ones mentioned above. Among these attacks, the four variants of the PGD attacks are used to be investigated in other application settings and are adapted to the adversarial malware detection domain for the first time.
The variant PGD attacks permit feature addition and feature removal, incurring larger manipulation sets than the Grosse, BGA, and BCA attacks. In these experiments, adversarial malware examples are generated by manipulating regular malware examples while preserving their malicious functionalities. Our empirical findings include: 
(i) standard DNNs without incorporating defense can be ruined by both grey-box and white-box attacks; (ii) adversarial regularization without considering attacks in the training phase has limited success in terms of improving the robustness of DNNs against adversarial examples; (iii) adversarial training with the Adam optimizer  can significantly enhance the robustness of DNNs against multiple grey-box evasion attacks, but not the more capable white-box Grosse, BCA and PGD-$\ell_1$ attacks; (iv) DAE provides a degree of extra robustness when used together with adversarial training, which is ineffective in defending against the white-box Grosse, BCA and PGD-$\ell_1$ attacks; (v) adding ensembles further improves the robustness of DNNs, at the price of sacrificing a degree of the effectiveness of adversarial training against the white-box PGD-$\ell_2$, PGD-$\ell_\infty$ and PGD-Adam attacks.

Third, we apply the framework to the AICS'2019 adversarial malware classification challenge organized by the MIT Lincoln Lab. According to the Challenge organizers, there were ``over 300 participants attempted to download and classify the malware data set'' \cite{aics_nes:Online} and we win the Challenge by achieving a 73.60\% Harmonic mean score (which is the metric the organizer chose to use before making the data available); i.e., we achieve the {\em highest} score among all of the participating teams.

Fourth, after announcing that we win the Challenge, the organizer makes the ground-truth labels publicly available at \url{http://www-personal.umich.edu/~arunesh/AICS2019/challenge.html}. In order to understand why we only achieve a 73.60\% Harmonic mean score, we leverage the 
ground-truth labels to conduct a further study. We show that (i) oversampling benefits adversarial regularization in defending against evasion attacks in term of the Macro-F1 score and (ii) adversarial regularization tends to overfit the perturbed examples while this phenomenon does not occur to the non-adversarial (i.e., original) data.

Fifth, we show that the framework is effective in resisting grey-box attacks via the widely-used Drebin Android malware dataset (with a 98.49\% accuracy on average), where the attacker knows some information about the defense and the defender knows some information about the attack.
When applied to the AICS'2019 challenge dataset but {\em only} considering the adversarial examples (for the sake of fair comparison with the experiment on the Drebin dataset), the framework only achieves a 76.02\% accuracy on average, where it is still true that neither the attacker knows the defense nor the defender knows the attacks.  This highlights that the defender should always strive to know as much information as possible about the attacks. In order to avoid any confusion, we reiterate that the aforementioned experiment result (i.e., 73.60\% in the Harmonic mean score) considers both adversarial and non-adversarial examples (as required by the challenge organizer);  whereas the 76.02\% accuracy disregards 
of the non-adversarial examples (for fair comparison with the experiment with the Drebin dataset). Another difference is that in the new experiment achieving a 76.02\% accuracy we use an ensemble of 5 building-block models, whereas in the experiment achieving 73.60\% Harmonic mean score we use 10 building-block models.

Last but not the least, we made our the code of our models publicly available at \url{https://github.com/deqangss/aics2019_challenge_adv_mal_defense}.

\subsection{Related Work} \label{sec:related-work}
Since the present paper focuses on defense against adversarial malware examples, we review related prior studies in 
four categories: {\em ensemble learning}, {\em input prepossessing}, {\em adversarial training/regularization}, and {\em DAE-based representation learning}.

Ensemble learning can reduce the generalization error by diversifying the building-block models. Biggio et al. \cite{biggio2010multiple,Biggio2010} show how the \emph{bagging} and \emph{random subspace} techniques can enhance the robustness of linear models against evasion attacks. Smutz and Stavrou \cite{smutz2016tree} propose using the confidence score produced by random forest classifiers to detect adversarial malware. Stokes et al. \cite{article_stokes} investigate the resilience of ensemble DNNs against evasion attacks. In this paper, we diversify the building-block models via randomly initialized parameters and the random subspace algorithm.

Input prepossessing transforms the input to a different representation, aiming to reduce the degree of perturbations applied to the original input.
For example, Random Feature Nullification (RFN) randomly nullifies features both in the training and test phases \cite{wang_2017}; HashTran \cite{li2018hashtran} reduces small perturbations using a locality-sensitive hashing; DroidEye \cite{YeFOSINT-SI-2018} quantizes binary representation via count featurization. In our framework, inspired by the idea of feature squeezing \cite{xu2017feature}, we use binarization to reduce the perturbations. 

Adversarial training augments the training data with adversarial examples. Various kinds of {\em heuristic} training strategies have been proposed (see, e.g., \cite{grosse2017adversarial,szegedyZSBEGF13,xu2014evasion,goodfellow6572explaining,kurakin2016adversarial}. However, these strategies typically deal with specific evasion methods and are not effective against others. Furthermore, researchers propose considering adversarial training with the optimal attack, which in a sense corresponds to the worst-case scenario and therefore could lead to classifiers that are robust against the non-optimal attacks \cite{madry2017towards,al2018adversarial}. In our framework, we seek the optimal attack via a gradient descent method, while meeting the requirement of discrete inputs via a nearest neighbor search.

Adversarial regularization is an adversarial training method that aims to train a model with {\em slightly} perturbed examples, which may or may not be  functionality-preserving. Intuitively, small perturbations benefit the generalization of DNN models \cite{drucker1992improving,Lyu:2015:UGR:2919336.2920639,miyato2016adversarial, goodfellow6572explaining,kurakin2016adversarial}.
This approach may be useful because in the context of malware detection, the defender may not know the manipulation set of the attacker.

DAE facilitates robust representation learning \cite{vincent2010stacked,mengc_2017}. Li et al. \cite{li2018hashtran} propose detecting adversarial malware examples using DAE. In our framework, we use DAE to learn the robust representation that is insensitive to perturbations.

\subsection{Paper Outline}

The rest of the paper is organized as follows. Section \ref{attack} presents the adversarial malware evasion attacks, including four attacks that are adapted to the domain of adversarial malware detection for the first time. 
Section \ref{method} describes our defense framework. 
Section \ref{validation} validates our defense framework with a real-world dataset.
Section \ref{exp} presents the results when applying the framework to the AICS'2019 Challenge {\em without} knowing anything about the attack.
Section \ref{exp-new} presents our further study after winning the AICS'2019 Challenge and being given the ground-truth labels of the test data. 
Section \ref{conclusion} concludes the present paper.

\section{Adversarial Malware Evasion Attacks} \label{attack}

\subsection{Notations}

Given a non-adversarial malware example $z \in \mathcal Z$, its feature representation $\mathbf{x} \in \mathcal{X}$ can be obtained via some {\em feature extraction} methods, where $\mathcal{Z}$ denotes the {\em example space} (i.e., the set of all possible software examples) and $\mathcal X$ denotes the {\em feature space}. A classifier $f:\mathcal{X} \rightarrow {\mathcal{Y}}$ takes ${\bf x}$ as input and outputs its label $y \in {\mathcal{Y}}$, where $\mathcal Y$ is the {\em label space}. 

We focus on a classifier $f$ that is learned as a neural network model ${\bf F}:\mathcal{X}\rightarrow\mathbb{R}^o$, whose output (\emph{softmax}) is treated as the probability mass function over $o=|\mathcal{Y}|$ classes \cite{al2018adversarial,grosse2017adversarial,aics_challenge:Online,DBLP:journals/corr/abs-1812-08108}. 
That is $f=\argmax_{j\in\mathcal{Y}}{\mathbf{F}}$, where $\argmax$ returns the index of the maximum element in a $o$-dimension vector. Let $L:\mathbb{R}^{o}\times\mathcal{Y}\to \mathbb{R}$ be a loss function. The parameters of $\mathbf{F}$, denoted by $\theta$, are optimized via
\begin{equation}
	\min_{\theta}\mathbb{E}_{(\mathbf{x},y)\in\mathcal{X}\times\mathcal{Y}}\left[L({\mathbf{F}(\mathbf{x})},y)\right]. \label{eq:nn-obj}
\end{equation}
Specifically, the cross-entropy is leveraged $L({\mathbf{F}(\mathbf{x})},y) = -\mathbbm{1}^\top_y\log({\mathbf{F}}(\mathbf{x}))$, where $\mathbbm{1}_y$ is the one-hot encoding vector for the label $y$.
For simplifying notations, we use $\mathbf{F}$ (rather than $\mathbf{F}_\theta$) to denote a neural network. Table \ref{table:notations} summarizes the main notations used in the paper.

\begin{table}[htbp!]
\centering\caption{Main notations used in the paper \label{table:notations}}
\begin{tabular}{l|p{.33\textwidth}}
\hline
Notation & Meaning\\\hline
$z\in\mathcal{Z}$ & $z$ is a software example; $\mathcal{Z}$ is the example space\\
$({\mathbf x}, y)\in \mathcal X \times \mathcal{Y}$ & ${\mathbf x}$ is feature representation of $z$; $\mathcal X$ is the feature space; $y$ is the label of ${\mathbf x}$; $\mathcal{Y}$ is the label space \\
$\delta_\mathbf{x}\in{\mathcal M}_\mathbf{x}$ & $\delta_\mathbf{x}$ is a perturbation vector of $\mathbf{x}$; ${\mathcal M}_\mathbf{x}$ is the manipulation set of $\mathbf{x}$ \\
${\mathbf x}'\in \mathcal S(\mathbf{x},\mathcal M_{\mathbf x})$ & ${\mathbf x}'$ is perturbed from ${\mathbf x}$; $\mathcal S$ is the set of perturbed representations derived from $\mathbf{x}$ and $\mathcal{M}_\mathbf{x}$; $\mathcal S\subseteq \mathcal X$\\
$o$ & $o$ is the number of classes \\
$dim$ & $dim$ is the number of dimension of $\mathbf{x}$ \\
$f:\mathcal{X}\to\mathcal{Y}$ & $f$ is the classifier \\
$\mathbf{F}:\mathcal{X}\to\mathbb{R}^o$ & $\mathbf{F}$ denotes a neural network \\
$\theta$ & $\theta$ denotes parameters of neural network $\mathbf{F}$ \\
$L:\mathbb{R}^o\times\mathcal{Y}\to\mathbb{R}$ & $L$ is cross-entropy loss function \\

\hline
\end{tabular}
\end{table}
\subsection{Basic Idea}

With regard to the feature space $\mathcal X$, given the representation-label pair $(\mathbf{x},y)$, the adversarial evasion attack attempts to perturb $\bf{x}$ into ${\bf{x'}}$, such that 
\begin{equation}
f({\mathbf x'}) \neq y 
~~\text{s.t.}~~\mathbf{x}' \in \mathcal{S}({\mathbf x},\mathcal{M}_{\mathbf x}) \label{n-ta} 
\end{equation}
where $\mathcal{S}({\mathbf x},\mathcal{M}_{\mathbf x})$ is the set of perturbed representations derived from the non-adversarial feature representation $\mathbf x$ and a \emph{manipulation set} $\mathcal{M}_{\mathbf x}$ (i.e. the set of manipulations that can preserve the malicious functionality of malware examples). The {\em perturbation vector} is denoted by $\delta_{\bf x}={\bf x'}-{\bf x}$ with $\delta_{\bf x}\in\mathcal{M}_\mathbf{x}$. Since the manipulation is conducted in the feature space, the attacker needs to map $\mathbf{x}'$ back into the example space $\mathcal Z$ in order to obtain an executable adversarial malware example $z' \in \mathcal Z$. This is a requirement that distinguishes adversarial malware detection from adversarial machine learning in other application domains, which induces the problem of generating adversarial examples in the discrete space. It is worth mentioning that 
an attacker tends to modify malware examples by exploiting one or multiple feasible manipulations \cite{7917369,grosse2017adversarial,li2020adversarial}.

\subsection{Threat Model} \label{sec:threat_model}

The threat model against malware classifiers and detectors can be specified by {\em what the attacker knows}, {\em what the attacker can do}, and {\em how the attacker wages the attack}.

\subsubsection{What the attacker knows} 
There are three kinds of models from this perspective.
A {\it black-box} attacker knows nothing about classifier $f$ except what is implied by $f$'s responses to the attacker's queries.
A {\it white-box} attacker knows all kinds of information about $f$, including its learning algorithms, model parameters, defenses strategies, etc.
A {\it grey-box} attacker knows an amount of information about $f$ that resides in between the preceding two extremes. For example, the attacker may know the feature set.

\subsubsection{What the attacker can do}
In evasion attack, the attacker only can manipulate malware examples in the test set using various kinds of manipulations, while obeying some constraints.
One constraint is to preserve the malicious functionality of malware. A simplifying assumption is to consider insertion only (e.g., flipping a feature value from `0' to `1' \cite{Biggio:Evasion,rndic_laskov,grosse2017adversarial,wang_2017,DBLP:journals/corr/RosenbergSRE17,Chen:2017:SES,al2018adversarial,pierazzi2020intriguing}, while noting that attackers can manipulate a malware example by inserting and deleting operations \cite{dang2017evading,anderson2017evading}. Since a larger manipulation set gives the attacker more freedom, we permit the attacker to conduct both {\em feature addition} and {\em feature removal}. The other constraint is to maintain the relation between features.  Using the AICS'2019 malware classification challenge as an example, we note that $n$-gram (uni-gram, bi-gram, and tri-gram) features reflect sequences of Windows system API calls. This means that when the attacker inserts an API call into a malware example, several features related to this API call will need to be changed according to the definition of $n$-gram features.

\subsubsection{How the attacker wages the attack} \label{sec:attack-mtds}

Researchers generate adversarial malware examples using various machine learning-based techniques such as genetic algorithms, reinforcement learning, and generative networks~\cite{316904628,anderson2017evading,Hu2017,xu2014evasion}. 
In order to generate adversarial malware examples effectively and efficiently, attackers often exploit the gradient-based methods~\cite{Biggio:Evasion,rndic_laskov,carliniW16a, goodfellow6572explaining,papernot_2016}. 
We here briefly describe multiple types of attacks, some of which were introduced in the context of malware detection but the others were introduced in the context of image classification and then adapted to the context of malware detection.

\medskip

\noindent{\bf Random Attack}. We introduce this method as a baseline attack in the adversarial malware detection domain. In this attack, the attacker randomly modifies a feature at each iteration until a predefined step is reached or no more features can be manipulated.

\medskip

\noindent{\bf Mimicry Attack}. This attack was introduced in \cite{Biggio:Evasion,rndic_laskov, 7917369,217486} for studying adversarial malware detection. In this attack, the attacker perturbs or manipulates a malware example such that the resulting adversarial version mimics a chosen benign example as much as possible. In order to reduce the degree of perturbations, the attacker may select the benign example to be close to the malware example that is to be modified.

\medskip

\noindent{\bf FGSM Attack}. This attack was introduced in the context of image classification \cite{goodfellow6572explaining} and then adapted to the malware detection  \cite{YeFOSINT-SI-2018,al2018adversarial}. FGSM perturbs a feature vector $\mathbf{x}$ in the direction of the $\ell_\infty$ norm of the gradients of the loss function with respect to the input, namely:
$$\mathbf{x}' = \Proj_{\mathcal{S}}\left(\mathbf{x} + \varepsilon\cdot \sign(\triangledown_\mathbf{x}L(\mathbf{F(\mathbf{x})}, y))\right),$$ 
where $\varepsilon>0$ is a scalar known as \emph{step size}, $\triangledown_\mathbf{x}$ indicates the derivative of the loss function $L(\mathbf{F(\mathbf{x})}, y)$ with respect to $\mathbf{x}$, and $\Proj_{\mathcal{S}}(\cdot)$ projects an input into $\mathcal{S}$ that denotes the shorthand of $\mathcal{S}(\mathbf{x},\mathcal{M}_\mathbf{x})$.

\medskip

\noindent{\bf Grosse Attack}. This attack was introduced by Grosse et al. \cite{grosse2017adversarial} in the context of malware detection. The attack considers sensitive features, namely the features have large positive gradients as far as the softmax output is concerned. The attack is to manipulate the absence of a feature (e.g., not making a certain API call) into the presence of the feature (i.e., making the API call), while preserving their malicious functionalities. These sensitive features can be identified by leveraging the gradients of the \emph{softmax} output of a malware example with respect to the input.

\medskip

\noindent{\bf BGA Attack} and \textbf{BCA Attack}. In the context of malware detection, Al-Dujaili et al. \cite{al2018adversarial} proposed two separate methods, dubbed BGA and BCA, aiming to solve: 
\begin{equation}
\max \limits_{\mathbf{x}' \in \mathcal{S}(\mathbf{x},\mathcal{M}_\mathbf{x})}L({\bf F}(\mathbf{x}'), y) \label{non_tar}.
\end{equation}
In addition, the authors considered malware examples in the binary space and restricted $\mathcal{M}_\mathbf{x}$ to API calls addition.
Both attack methods iterate multiple steps. In each iteration, BGA increases the feature value from `0' to `1' if the corresponding partial derivative of the loss function with respect to the input is greater than or equal to the gradient's $\ell_2$ norm divided by $\sqrt{dim}$, where $dim$ is the input dimension. In contrast, BCA flips `0' to `1' for a component at the iteration corresponding to the maximum gradient of the loss function with respect to the input. 

\medskip

\noindent{\bf PGD Attack}. We adapt the PGD method to the context of malware detection, by accommodating discrete input spaces. In contrast to the Grosse, BGA, and BCA attacks, the adapted PGD attacks permit both feature addition and feature removal. Specifically, PGD finds perturbations via an iterative procedure

\begin{equation}
\delta^{i+1}_\mathbf{x}=\Proj_{\hat{\mathcal{M}}_\mathbf{x}}\left(\delta^i_\mathbf{x}+\alpha\cdot\triangledown_{\delta_\mathbf{x}}L(\mathbf{F}(\mathbf{x} + \delta^i_\mathbf{x}), y)\right), \label{eq:pgd}
\end{equation}
where $\alpha>0$ is the step size, $\triangledown_{\delta_\mathbf{x}}$ is the derivative of the loss function $L(\mathbf{F}(\mathbf{x} + \delta^i_\mathbf{x}), y)$ with respect to $\delta_\mathbf{x}$, and $\Proj_{\mathcal{\hat{M}}_\mathbf{x}}$ projects perturbations into a predetermined space $\mathcal{\hat{M}}_\mathbf{x}$. We set $\mathcal{\hat{M}}_\mathbf{x}=[\underline{\mathbf{u}},\overline{\mathbf{u}}]$ for $\underline{\mathbf{u}} = \minimum(\mathcal{M}_\mathbf{x})$ and $\overline{\mathbf{u}}= \maximum(\mathcal{M}_\mathbf{x})$, where $\minimum$ returns the component-wise minimum vector (i.e., each component of the vector corresponding to the minimum of the corresponding component values of the vectors in $\mathcal{M}_\mathbf{x}$) and $\maximum$ returns the component-wise maximum vector.

When solving Eq.\eqref{eq:pgd}, we encounter two issues that need to be addressed: (i) small derivatives of $\mathbf{g}=\triangledown_{\delta_\mathbf{x}}L$ and (ii) mapping perturbations into discrete space $\mathcal{M}_\mathbf{x}$. 
To see issue (i), we note that by writing $\mathbf{F}$ as $\mathbf{F}(\mathbf{x})=\softmax(\mathbf{Z}(\mathbf{x}))$, we have ${\partial L}/{\partial\delta_\mathbf{x}}=(\mathbf{F}-\mathbbm{1}_y)\cdot{\partial \mathbf{Z}}/{\partial\delta_\mathbf{x}}$, meaning that the derivatives approach zero when $\mathbf{F}$ predicts $\mathbf{x}$ as $y$ with high confidence. To cope with this, researchers \cite{madry2017towards,DBLP:journals/corr/abs-1904-13000} have proposed to ``normalize'' the derivatives using $\ell_p$-norm and leveraging the steepest direction as follows:
\begin{itemize}
\item For $p=1$, the direction is $\sign(g_i)\mathbf{1}_i$, where $i$ is the index corresponding to the maximum absolute value of $\mathbf{g}=(g_1,\ldots,g_{dim})$ with $dim$ being the number of input dimension, $\mathbf{1}_i$ has the same dimension as $\mathbf{g}$ and has value $1$ at the $i$th component and value $0$ at the other components, and $\sign$ returns $1$ when the input $>0$, $-1$ when the input $<0$, and $0$ when the input $=0$.
\item For $p=2$, the direction is $\mathbf{g}/\parallel \mathbf{g} \parallel_2$.
\item For $p=\infty$, the direction is sign of gradients, i.e., $\sign(\mathbf{g})$.
\end{itemize}
We call these variant PGD attacks PGD-$\ell_1$, PGD-$\ell_2$ and PGD-$\ell_\infty$, respectively. Note that PGD-$\ell_1$ degrades to the BCA attack when only feature addition is permitted. In addition to these $\ell_p$-norm based attacks, we observe that the attacker can use the Adam optimizer to accelerate the process of gradient descent (the ``normalized'' gradients are approximate to $\pm 1$) \cite{kingmaB14}, leading to a new variant of the PGD attack, which we call PGD-Adam.

\begin{algorithm}[!htbp]
	\KwIn{The feature representation-label pair $(\mathbf{x},y)$, manipulation set $\mathcal{M}_\mathbf{x}$, number of iterations $T$, step size $\alpha$}
	\KwOut{Perturbed example $\mathbf{x}'$}
	
	Initialize perturbation vector $\delta_\mathbf{x}^0=\mathbf{0}$;
	
	Derive the continuous space $\hat{\mathcal{M}}_\mathbf{x}$ and the perturbed representation set $\mathcal{S}$;
	
	\For{$i=0$ to $T-1$}{
		Obtain the derivatives $\triangledown_{\delta_{\mathbf{x}}}L$ and normalize them using $\ell_p$-norm where $p=1,2,\infty$ or the Adam method;
		
		Calculate $\delta^{i+1}_{\mathbf{x}}$ via the Eq.\eqref{eq:pgd};
	}
	
	Obtain $\mathbf{x}'$ by mapping $\tilde{\mathbf{x}}'=\mathbf{x}+\delta^{T}_{\mathbf{x}}$ via Eq.\eqref{eq:map}; 
	
	\Return $\mathbf{x}'$.
	
	\caption{PGD attack in the feature space.}
	\label{alg:pgd_attack}
\end{algorithm}

To address the issue (ii), we introduce a mapping method to consider two settings as follows. In order to follow the direction of ``normalized'' gradients, let the perturbation $\delta_\mathbf{x}$ be continuous during the optimization process. We map the perturbed representation obtained at the last iteration, denoted by $\tilde{\mathbf{x}}'=(\tilde{x}'_1,\ldots,\tilde{x}'_{dim})$, into $\mathcal{S}$ by selecting its closest neighbor $\mathbf{x}'=(x'_1,\ldots,x'_{dim})$ such that
\begin{equation}
\mathbf{x}'=\argmin_{\mathbf{x}'\in\mathcal{S}}\parallel\mathbf{x}'-\tilde{\mathbf{x}}'\parallel_1=\argmin_{\mathbf{x}'\in\mathcal{S}}\sum_{i=1}^{dim}|x'_i-\tilde{x}'_i|. \label{eq:map}
\end{equation}
Geometrically speaking, Eq.\eqref{eq:map} says that for the $i$th dimension, $x'_i$ is the feasible scalar closest to $\tilde{x}'_i$.
Algorithm \ref{alg:pgd_attack} summarizes the PGD attacks in the feature space.

\section{Adversarial Malware Defense}\label{method}

\subsection{Guiding Principles} \label{sec:principles}

These principles are geared to neural network classifiers, which are chosen as our focus because deep learning techniques are increasingly employed in the domain of malware detection/classification, but their vulnerability to adversarial evasion attack has yet to be tackled~\cite{raff2017malware}.

\subsubsection{Principle 1: Knowing the enemy}

This principle says that the defender should strive to extract useful information about the attacks as much as possible as the information will offer insights on designing countermeasures. Threat models are a standard approach to modeling attacks. Moreover, it is possible to design some indicators of adversarial examples.
On the other hand, the attack method and manipulation set may not be known to the defender. This means that whenever possible, the defender has to simulate them.

\subsubsection{Principle 2: Bridging grey-box vs. white-box gap}

In grey-box attacks, the attacker knows some information about the feature set and therefore can train a surrogate classifier $\hat{f}:\mathcal{X}  \rightarrow \mathcal{Y}$ from a training set (where the realization of $\hat{f}$ is a neural network $\hat {\bf F}$), leveraging the transferability from $\hat{f}$ to $f$ to generate adversarial examples. 
Consider an input ${\bf x}$ for which a grey-box attacker generates perturbations using
\begin{equation}
    \hat \delta_{\bf x} = \argmax \limits_{||\delta_{\bf x}||\leq\eta~\land~\delta_{\bf x} \in \mathcal{M}_{\bf x}}L({\bf \hat F}({\bf x + \delta_x}), y), \nonumber
\end{equation}
where $\eta$ is an upper bound and possibly large. Based on the degree of perturbations, we consider two cases: (i) $\eta$ is small and (ii) $\eta$ is large. We further assume that the optimal perturbation vector $\delta_{\mathbf x}$ of $\mathbf{F}$ exists.
\ignore{
\begin{assumption}
We assume that the optimal perturbation to $\mathbf{F}$ exists, that is $\delta_{\mathbf x}=\argmax_{\delta_{\mathbf x}\in \mathcal{M}}(L(\mathbf{F}(\mathbf{x}+\delta_{\mathbf x}),y)$.
\end{assumption}
}

In case (i) or when $\eta$ is small, the change to the loss of $f$ incurred by $\hat \delta_{\bf x}$ is
\begin{align*}
	\left|\Delta L\right| & =\left|L(\mathbf{F}(\mathbf{x} + \hat \delta_{\mathbf x}), y) - L(\mathbf{F}(\mathbf{x}), y)\right| \\
	& \approx \left|\triangledown L(\mathbf{F}(\mathbf{x}), y)^\top \hat \delta_{\mathbf x}\right|
	\leq~\max_{||\delta||\leq\eta}\left|\triangledown L(\mathbf{F}(\mathbf{x}), y)^\top \delta\right| \\
	& = \eta||\triangledown L||_*,
\end{align*}
where the approximation is derived using the first-order Taylor expansion at point $\mathbf{x}$, $\triangledown$ is the operator for computing partial derivatives of the loss function with respect to the input of neural network $\mathbf{F}$, and ``$||\cdot||_*$'' is the dual norm of $||\cdot||$.

In case (ii) or when $\eta$ is large, we derive
\begin{align*}
    \left|\Delta L\right| &= \left|L({\bf F}({\bf x} + \hat \delta_{\bf x}), y) - L({\bf F}({\bf x}), y)\right|\\
    &=\left|\int_{0}^{\hat \delta_{\bf x}}{\triangledown L({\bf F}({\bf x}+\delta), y)}d\delta\right|\\
    &=\left|\int_{0}^{1}\triangledown L({\bf F}({\bf x}+t\hat \delta_{\bf x}), y)^\top \hat \delta_{\bf x}dt\right|\\
    &\leq\eta\sup \limits_{||\delta||\leq\eta}\left\|\triangledown L({\bf F}({\bf x} + \delta), y)\right\|_*.
\end{align*}
The preceding observation shows that corresponding to the same perturbation upper bound $\eta$, the loss incurred by grey-box attacks is upper bounded by the loss incurred by white-box attacks. This suggests us to focus on the robustness of classifier $f$ against the optimal white-box attack.

\subsubsection{Principle 3: Not putting all eggs in one basket}

This is suggested by the observation that no single classifier may be effective against all kinds of attacks. An ensemble can be built by many methods (e.g., bagging, boosting, or stacking) \cite{zhou2012ensemble}. For example, {\it random subspace}~\cite{709601_rss} is seemingly particularly suitable for formulating malware classifier ensembles owing to the high dimensional feature vector of malware, which indicates a high vulnerability of classifiers to adversarial malware examples~\cite{7917369,demontis2019adversarial}. 

Formally, an ensemble $f_{en}:\mathcal{X}\rightarrow\mathcal{Y}$ contains a set of neural networks $\{\mathbf{F}_i\}_{i=1}^l$, namely $ \{\mathbf{F}_i:\mathcal{X}\rightarrow\mathbb{R}^o\}$ for $1\leq i \leq l$.  
Given a test example ${\bf x}$, we treat the base model equally, as suggested by the study \cite{biggio2010multiple,DBLP:journals/corr/abs-1811-09300}, and the voting method is 
\[
    \mathbf{F}_{en}(\mathbf{x})= \frac{1}{l} \sum_{i=1}^{l}\mathbf{F}_i(\mathbf{x}).
\]
We obtain the predicted label by $f_{en}=\argmax_{j\in\mathcal{Y}}\mathbf{F}_{en}$.

\subsubsection{Principle 4: Using transformation against perturbation}

In typical applications, the defender does not know what kinds of evasion attacks are waged by the attacker. These attacks can produce a spectrum of perturbations, from manipulating a few features (e.g., the PGD-$\ell_1$ attack) to manipulating a large number of features (e.g., the FGSM attack). Moreover, we may give higher weights to the transformation techniques that can simultaneously reduce the degrees of multiply types of perturbations such as $\ell_{\infty}$ norm, $\ell_1$ norm, or $\ell_2$ norm. This suggests us to use the {\it binarization} technique~\cite{xu2017feature,schott2018towards}:  
When the feature value of the $i$th feature, denoted by $x_i$, is smaller than a threshold $\Theta_i$, we binarize $x_i$ to 0; otherwise, we binarize $x_i$ to 1.

\subsubsection{Principle 5: Using vaccine}

We harden a model incorporating the known {\em minmax} adversarial training:
\begin{equation}
    \min \limits_{\theta}~\mathbb{E}_{({\bf x}, y)\in \mathcal{X}\times\mathcal{Y}}\left[ L(\mathbf{F}(\mathbf{x}),y) + \max \limits_{{\mathbf x}' \in \mathcal{S}}L({\bf F}({\bf x}'), y)\right] \label{loss:clf}.
\end{equation}
Al-Dujaili et al. \cite{al2018adversarial} instantiate this method by using attacks with feature addition solely (e.g., BGA). 
In order to accommodate more manipulations, we solve the problem of inner maximization using the Adam optimizer (see Section \ref{sec:attack-mtds}). Given the issue of local minima, we run the inner maximizer several times, each with a random initial point near the training data, and then select the point that maximizes the loss function of $L$.

It is worth mentioning that in the AICS'2019 Challenge, the defender does not know the manipulation set $\mathcal{M}_\mathbf{x}$ and thus cannot derive $\mathcal{S}$. In this case, we propose training malware classifiers by applying  small perturbations to the feature representations of malware examples ({\em without} necessarily preserving their malicious functionalities). This would benefit model generalization \cite{drucker1992improving,goodfellow6572explaining}. Let a norm $||\cdot||$ measure the perturbation $\delta_{\mathbf x}$ with $\eta$ bounded. We have $\max_{||\delta_{\mathbf x}|| \leq \eta}L({\bf F}(\mathbf{x} + \delta_\mathbf{x}), y) \approx
L({\bf F}({\bf x}), y) + \eta||\triangledown L(\mathbf F(\mathbf{x}),y)||_*$, leading to $|L({\bf F}(\mathbf{x}'), y) - L({\bf F}(\mathbf{x}), y)| \leq \eta||\triangledown L(\mathbf F(\mathbf{x}),y)||_*$. Therefore, adversarial regularization assures that small perturbations do not change the prediction significantly.

\subsubsection{Principle 6: Preserving semantics}

This suggests us to strive to learn neural network models that are sensitive to malware semantics, but not the perturbations because adversarial examples must retain the malicious functionality of original malware. Specifically, we propose using denoising auto-encoder to learn semantics-preserving representations, rendering neural network less sensitive to perturbations. A DAE $ae=d \circ e$ unifies two components: an encoder $e:\mathcal{X}\rightarrow{\mathcal{H}}$ that maps an input $M({\bf x})$ to a latent representation ${\bf r} \in \mathcal{H}$ and a decoder $d: \mathcal{H}\rightarrow{\mathcal{X}}$ that reconstructs ${\bf x}$ from ${\bf r}$, where the $\mathcal{H}$ is the latent representation space and $M$ refers to some operations applied to ${\bf x}$ (e.g., adding Gaussian noises to ${\bf x}$). Vincent et al. \cite{vincent2010stacked} showed that the lower bound of the {\it mutual information} between ${\bf x}$ and ${\bf r}$ is maximized when the reconstruction error is minimized. In the case of Gaussian noise $\epsilon \sim \mathcal{N}(0, \sigma^2)$ and reconstruction loss
\begin{equation}
    \mathbb{E}_{\epsilon \sim \mathcal{N}(0, \sigma^2)}\left\lVert ae({\bf x} + \epsilon) - {\bf x})\right\rVert_2^2,
\end{equation}
Alain and Bengio \cite{alain2014regularized} showed that the optimal $ae^*({\bf x})$ is
\begin{equation}
    ae^*({\bf x})=\frac{\mathbb{E}_\epsilon \left[p({\bf x} - \epsilon)({\bf x} - \epsilon)\right]}{\mathbb{E}_\epsilon \left[p({\bf x} - \epsilon)\right]} \label{eq:opt-ae},
\end{equation}
where $p(\cdot)$ is the probability density function. Eq.\eqref{eq:opt-ae} says that representations of a well-trained DAE are insensitive to ${\bf x}$ because of the weighted average from the neighborhood of ${\bf x}$, which is reminiscent of the {\em attention} mechanism~\cite{luong2015effective}. This means that DAE can handle certain types of small perturbations.
To learn a DAE model, we leverage two kinds of noise: (i) {\it Salt-and-pepper noise $\epsilon$}: A small fraction of elements of original example ${\bf x}$ are randomly selected, and then set their values as their respective minimum or maximum.
(ii) {\it Adversarial perturbation $\delta_{\bf x}$}: A perturbation $\delta_{\bf x}$ is added to ${\bf x}$ such that classifier $f$ or base classifier $f_{i}$ misclassifies ${\bf x'}={\bf x} + \delta_{\bf x}$.
Given a training example $\mathbf{x}$ over the feature space $\mathcal{X}$,
the risk of a denoising auto-encoder is
\begin{equation}
    \min \limits_{\tilde\theta,\xi} \mathbb E_{\mathbf{x} \in \mathcal X} \left[L_{ae}(\mathbf{x}, ae(\mathbf{x} + \epsilon)) + L_{ae}(\mathbf{x}, ae({\mathbf x}'))\right] \label{loss:dae},
\end{equation}
\noindent
where $L_{ae}:\mathcal X \times \mathcal X \mapsto \mathbb{R}$ calculates the mean-square error, the learnable parameters $\tilde\theta$ and $\xi$ respectively belongs to the encoder and decoder.

\begin{figure*}[!htbp]
	\centering
	\scalebox{0.4}{
	\includegraphics{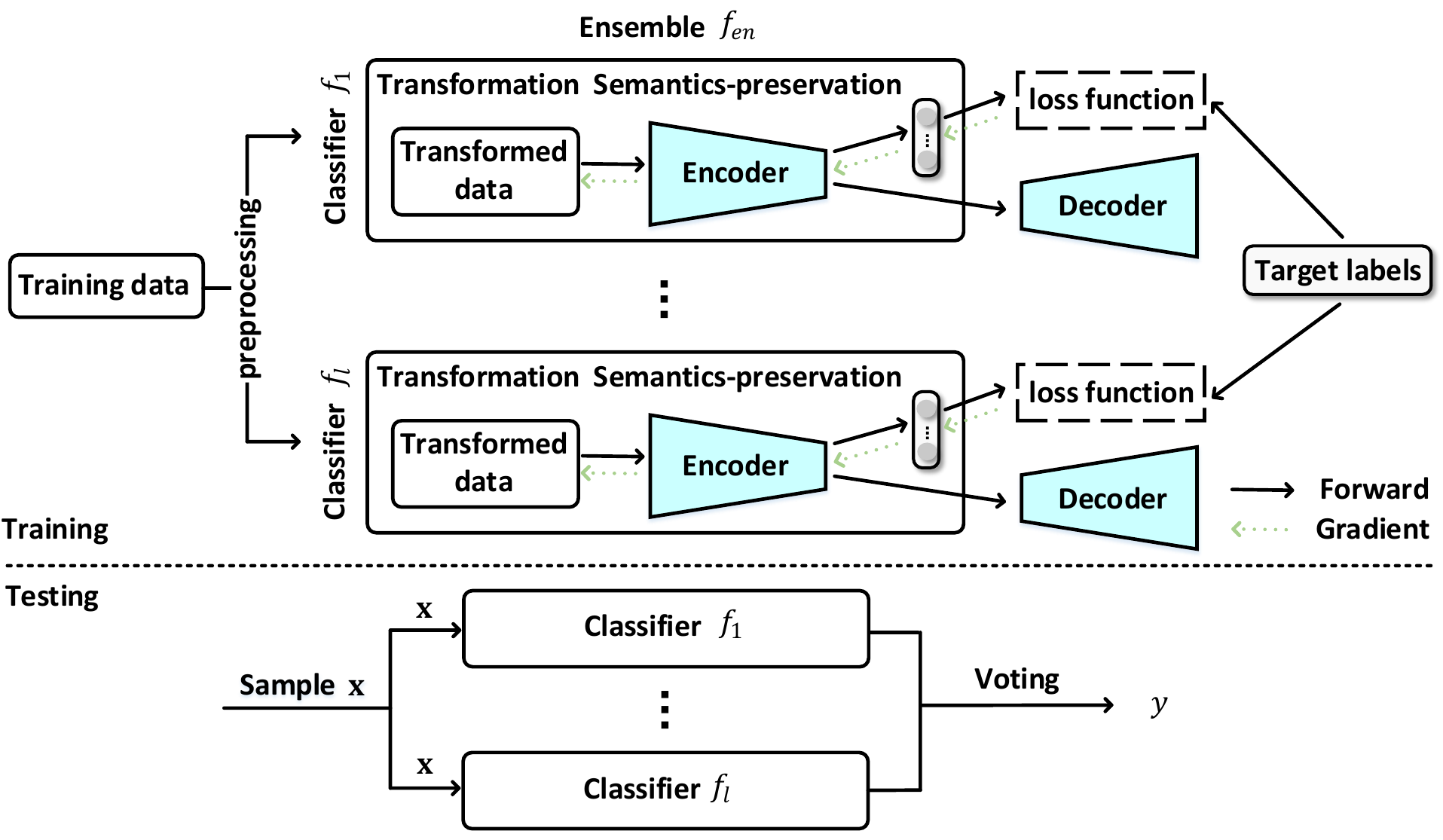}
	}
\caption{Overview of the adversarial malware defense framework. In the training phase, an ensemble of $l$ neural network classifiers are trained, with each classifier hardened by three countermeasures (i.e., input transformation, semantics-preserving, and adversarial training on the transformed data). 
}
	\label{fig:wf}
\end{figure*}

\subsection{Turning Principles into A Framework}

The principles discussed above guide us to propose a framework for adversarial malware classification and detection, which is highlighted in Figure \ref{fig:wf} and elaborated below. Specifically, we first examine whether the attacks have some useful information that could to be incorporated via a proper preprocessing (according to Principle 1). 
We propose using an ensemble $f_{en}$ of classifiers $\{f_i\}_{i=1}^l$ (according to Principle 3), which are trained from random subspace of the original feature space. Each classifier $f_i$ is hardened by three countermeasures: input transformation via binarization (according to Principle 4); adversarial training/regularization models on the attacks using Adam optimizer (dot arrows in Figure \ref{fig:wf}, according to Principle 2 and 5); semantics-preservation is achieved via an encoder and a decoder (according to Principle 6). In order to attain adversarial training and at the same time semantics-preservation, we learn classifier $f_i$ via block coordinate descent to optimize different components of the model. 

\begin{algorithm}[!htbp]
\KwIn{The training set $(X, Y)$, number of repeat times $K$, epoch $N_{epoch}$ and mini-batch size $N$.}


Select a ratio $\Lambda$ of sub-features from the feature set;

Transform input $X$ to $\overline{X}$ via binarization;

\For{$epoch=$ \rm ${1}$ to $N_{epoch}$}{
Sample a mini-batch $\{\mathbf{x}_i, y_i\}_{i=1}^N$ from the $(\overline{X}, Y)$;
\BlankLine

\For{$k=$ \rm ${1}$ to $K$}{
Apply slight salt-and-pepper noises to $\{\mathbf{x}_i\}_{i=1}^N$;

Derive the perturbed representation $\{{\mathbf{x}'}_i^{k}\}_{i=1}^N$ via Algorithm \ref{alg:pgd_attack} using Adam method;

}

Select $\mathbf{x}'_i$ from $\{{\mathbf{x}'}_i^{k}\}_{k=1}^{K}$ for $\mathbf{x}_i$ $(i=1,\cdots,N)$ to maximize the cross-entropy loss;

\BlankLine

Calculate the reconstruction loss via Eq.\eqref{loss:dae};

Backpropagate the loss and update the denoising autoencoder parameters;

\BlankLine

Calculate the adversarial training loss via Eq.\eqref{loss:clf};

Backpropagate the loss and update classifier parameters;
}

\caption{Training classifier $f_i$}
\label{alg:train}
\end{algorithm}

Algorithm \ref{alg:train} integrates all pieces for training individual classifiers. The training procedure consists of the following steps. (i) Given a training set $(X, Y)$, we randomly select a ratio $\Lambda$ of sub-features to the feature set, and then transform $X$ into ${\overline{X}}$ via the binarization discussed above. (ii) We sample a mini-batch $\{{\bf x}_i, y_i\}_{i=1}^N$ from $(\overline{X}, Y)$, and calculate adversarial examples ${\bf x}'_i$ for ${\bf x}_i \in \{{\bf x}_i\}_{i=1}^N$ according to Lines 5-9 in Algorithm \ref{alg:train}. (iii) We pass the $\{{\bf x}'_i\}_{i=1}^N$ through the denoising auto-encoder to compute the reconstruction loss with respect to the target $\{{\bf x}_i\}_{i=1}^N$ via Eq.\eqref{loss:dae}, and update the parameters of the denoising auto-encoder. (iv) We pass both the $\{{\bf x}_i + \delta_{{\mathbf x}_i}\}_{i=1}^N$ and $\{{\bf x}_i\}_{i=1}^N$ through the neural networks to compute the classification error with respect to the ground-truth label $\{y_i\}_{i=1}^N$ via Eq.\eqref{loss:clf}, and update the parameters of the classifier via backpropagation. Note that Steps (ii)-(iv) are performed in a loop. The output of the training algorithm is a neural network classifier.

\section{Validating Framework via Drebin Dataset}\label{validation}

We validate the effectiveness of the framework using the Drebin dataset of Android malware \cite{Daniel:NDSS}, while considering 11 grey-box attacks and 9 white-box attacks. This dataset also applied by former studies in the domain of adversarial malware detection \cite{7917369,grosse2017adversarial,li2018hashtran,236234}.

\subsection{Data Pre-Processing}

\textbf{Dataset.} 
The Drebin dataset \cite{Daniel:NDSS} contains 5,615 malicious Android packages (APKs), and also provides features of 123,453 benign examples, together with their SHA256 values but not the examples themselves. All samples were labeled using the VirusTotal service \cite{VirusTotal:Online} before the year 2015. An example was treated as malicious if there are at least two scanners say it is malicious, and is treated as benign if no scanners detect it \cite{Daniel:NDSS}. Because the VirusTotal may update the detection result along with the time \cite{10.1007/978-3-319-45719-2_11}, we consider relabeling the APKs. We download benign applications corresponding to the given SHA256 values from the APK markets (e.g., Google Play, AppChina, etc.), and collect 54,829 APKs in total. We send all of these examples (i.e., malicious and benign alike) to the VirusTotal service again.
Surprisingly, 12,496 benign APKs are detected as malicious (rather than benign) by at least one scanners, and most of them are detected as \emph{Adware} or \emph{Trojan}; this suggests that the original Drebin training set has been contaminated by the \emph{poisoning attack}. This may be caused by some of the following reasons:
(i) Virustotal updates the scanners over the time; (ii) Virustotal updates the report of a file when a user requires to rescan the file; (iii) after an update, the previous report cannot be obtained anymore. We thus remove these 12,496 benign examples from the original benign dataset, leaving 42,333 benign APKs.
The resulting dataset contains 5,615 malicious APKs and 42,333 benign APKs, namely 47,948 examples in total. We randomly split the dataset into three disjoint sets for training (60\%), validation (20\%), and test (20\%), respectively.

\smallskip

\noindent\textbf{Feature Extraction.} APK is an archive file containing \emph{AndroidManifest.xml}, \emph{classes.dex}, and others (e.g., \emph{res}, \emph{assets}). The \emph{manifest} file describes an APK's information, such as the name, version, announcement, library files used by the application. The source code is compiled to build the \emph{.dex} file which is understandable by the Java Virtual Machine. 
The Drebin dataset has eight subsets of features, including four subsets of features extracted from {\em AndroidManifest.xml} (denoted by $S_1, S_2, S_3, S_4$, respectively) and four subsets of features extracted from the disassembled dexcode (denoted by $S_5, S_6, S_7, S_8$, respectively). More specifically, ({\bf i}) $S_1$ contains features corresponding to the access of an APK to the hardware of a smartphone (e.g., camera, touchscreen, or GPS module); ({\bf ii}) $S_2$ contains features corresponding to the permissions requested by the APK in question;
({\bf iii}) $S_3$ contains features corresponding to the application components (e.g., {\em activities}, {\em service}, {\em receivers}, etc.); ({\bf iv}) $S_4$ contains features corresponding to the APK's communications with the operating system; ({\bf v}) $S_5$ contains features corresponding to the critical system API calls, which cannot run without appropriate permissions or the {\em root} privilege; ({\bf vi}) $S_6$ contains features corresponding to the used permissions; ({\bf vii}) $S_7$ contains features corresponding to the API calls that can access sensitive data or resources on a smartphone;  ({\bf viii}) $S_8$ contains features corresponding to IP addresses, hostnames and URLs found in the disassembled code. 

In order to extract applications' features, we use Androgurad 3.3.5, a reverse engineering toolkit for APK analysis \cite{Androguard:Online}. Note that 141 APKs cannot be analyzed. Moreover, a feature selection is conducted to  remove those low-frequency features for the sake of computational efficiency. As a result, we keep 10,000 features with high frequencies. The APK is mapped on the feature space as a binary feature vector, where `1' (`0') represents the presence (absence) of a feature in the APK. 

\subsection{Training Classifiers}

\textbf{Classifiers.} In order to validate the defense framework, we use and compare five classifiers: (\textbf{i}) the basic DNN with no effort made to defend adversarial examples; (\textbf{ii}) hardened DNN incorporating adversarial training with known manipulation set (dubbed Adversarial Training), which manifests Principle 2 (grey-box attacks can be bounded by the worst-case white-box attack) and Principle 5 (min-max adversarial training); (\textbf{iii}) hardened DNN incorporating adversarial regularization because the defender may know nothing about the manipulation set, which is true in the case of AICS'2019 Challenge (dubbed Adversarial Regularization); (\textbf{iv}) Denoising Auto-Encoder (DAE) based-classifier, which manifests Principle 6 (semantics-preserving representations); (\textbf{v}) classifier hardened by both Adversarial Training and DAE (dubbed AT+DAE); (\textbf{vi}) ensemble of AT+DAE classifiers in the random subspace (manifesting Principle 3, dubbed Ensemble AT+DAE). For Principle 1 (i.e., knowing your enemy), we will simulate attacks in the next subsection. Since we use binary feature vector, Principle 2 (binarization) is not applicable.

\noindent
\textbf{Hyper-parameters Setting.} We use DNNs with two fully-connected hidden layers (each layer having 160 neurons) with ReLU activation function. All classifiers are optimized by using Adam with epochs 150, mini-batch size 128, and learning rate 0.001. For Adversarial Training, the inner maximization is optimized by using Adam with learning rate 0.02 and iteration steps $T = 100$ to search adversarial examples as many as possible. For Adversarial Regularization, we set the learning rate as 0.01 for Adam and conduct preliminary experiments to tune a proper iteration step $T$. Finally, we set $T=60$. We use an ensemble of 5 base classifiers. Our preliminary experiments suggest us to learn base classifiers from the entire training set and the entire feature set. Unless with special mentioning, all classifiers that require to solve the inner maximization are trained without random starting points so as to ease the analysis (i.e., $K=0$).


\subsection{Attack Experiments and Classification Results} 

We present threat models specified by whether the attacker wages grey-box or white-box attacks, and constraints on the attacker's manipulation set.

\noindent
\textbf{Grey-box vs. White-box Attacks.}
We consider two attack scenarios. (\textbf{i}) \emph{Grey-box attacks}: In this setting, we simulate the attack scenario of the AICS'2019 Challenge organizers. That is, the attacker knows the dataset, feature set, but not the defender's learning algorithm. The attacker generates adversarial examples from a surrogate classifier. We consider a surrogate model of two fully-connected hidden layers (200 neurons each layer) and learn the model on the training set. (\textbf{ii}) \emph{White-box attacks}: In this setting, the attacker knows everything about the malware detector. Therefore, the adversarial examples are directly generated from the corresponding malware detector.

\noindent
\textbf{Manipulations Constraints.} 
Given an APK, we consider both \emph{incremental} and \emph{decremental} manipulations. The incremental manipulation allows the attacker to insert some objects (e.g., \emph{activity}) into an APK example to avoid detection. The decremental manipulation allows the attacker to hide some objects (e.g., \emph{activity}) to avoid detection. In any case, the adversarial example should preserve the malicious functionality of the malware from which the adversarial example is generated.

When the attacker uses incremental manipulations, the attacker can insert some manifest features (e.g., request extra permissions and hardware, state additional \emph{services}, \emph{Intent-filter}, etc.). However, some elements are hard to insert, such as \emph{content-provider}, because the absence of URI will corrupt an APK example. With respect to the \emph{.dex} file, a dead function or class (which is never called) containing specified system APIs can be injected without destroying the APK example. The similar means can be performed for the \emph{string} injection (e.g., IP address), as well.

\lstset{
	frame = lrbt,
	numbers = left,
	aboveskip=3mm,
	belowskip=3mm,
	showstringspaces=false,
	columns=flexible,
	basicstyle={\small\ttfamily},
	language=java,
	numbers=none,
	numberstyle=\tiny\color{gray},
	keywordstyle=\color{blue},
	commentstyle=\color{dkgreen},
	stringstyle=\color{mauve},
	breaklines=true,
	breakatwhitespace=true,
	tabsize=3
}

\begin{lstlisting}[caption=Java code to hide the API method ``println''.,
label={lst:java-example}, captionpos=b]
public void hideAPI() throws Exception{
// hide 'println'
String e_str = "ExMLXEZUDw";
// get 'println'
String p_str = decryptStr(e_str);
Class c = java.lang.System.class;
Field f = c.getField("out");
Method m = f.getType().
getMethod(p_str,String.class);
m.invoke(f.get(null), "hello world!");
return void
}
\end{lstlisting}

When the attacker uses decremental manipulations, the APK's information in the \emph{xml} files can be changed (e.g., package name). However, it is impossible to remove \emph{activity} entirely because an \emph{activity} may represent a class implemented in the \emph{.dex} code. Nevertheless, we can rename an \emph{activity} and change its relevant information (e.g., \emph{activity label}), while noting that the related components in the \emph{.dex} should be modified accordingly. The other components (e.g., \emph{service}, \emph{provider} and \emph{receiver}) also can be modified in the similar fashion, and the resource files (e.g., images, icons) can be manipulated as well. In terms of \emph{dexcode}, the method names and class names that are defined by developers could be modified, too. Note that the corresponding statement, instantiation, reference, and announcements should be changed accordingly. Moreover, user-specified \emph{strings} can be obfuscated using encryption and the cipher-text will be decrypted at running time. Further, the attacker can hide \emph{public} and \emph{static} system APIs using Java reflection and encryption together. This is shown by the example in List \ref{lst:java-example}. All of the modifications mentioned above only obfuscate an APK without changing its functionalities.

One challenge is that the attacker needs to perform fine-grained manipulations on compiled files automatically at scale, while preserving the functionalities of malware examples. This important because a small change in a malware example can render the file unexecutable. Since Android APIs have upgraded multiple times in the past 5 years, the attacker has to inject compatible APIs into an APK when manipulating a malware example. The preservation of malicious functionalities may be estimated by using a dynamic malware analysis tool, (e.g., Sandbox).

\smallskip
\noindent
\textbf{Mapping Manipulations to Feature Space.} The aforementioned manipulations modify static Android features, such as API calls and components in the manifest file. Two kinds of perturbations can be applied to the Drebin feature space. (i) {\em Feature addition.} The attacker can increase the feature values (e.g., flipping `0' to `1') of appropriate objects, such as components (e.g., \emph{activity}), system APIs, and IP address.
(ii) {\em Feature removal.} The attacker can flip `1' to `0' by removing or hiding objects (e.g., \emph{activity}, APIs.)
Table \ref{tab:map_feature_sets} summarizes our manipulations in the Drebin feature space. We observe that neither feature addition nor feature removal can be applied to $S_6$ because these features depend on $S_2$ and $S_5$, meaning that modifications on $S_2$ or $S_5$ may cause changes to features in $S_6$.

\begin{table}[!htbp]
\caption{Overview of manipulations on feature space, where \checkmark (\xmark) indicates that the feature addition or removal operation can (cannot) be performed on features in the corresponding subset.}
  \centering
\begin{tabular}{ l|l|c|c }
  \toprule
  \multicolumn{2}{ c |}{\textbf{Feature sets}} & Addition & Removal \\
  \midrule
  \multirow{4}{*}{\texttt{manifest}}
    & $S_{1}$ Hardware & \checkmark & \xmark \\
    & $S_{2}$ Requested permissions & \checkmark & \xmark \\
    & $S_{3}$ Application components & \checkmark & \checkmark \\
    & $S_{4}$ Intents & \checkmark & \xmark \\
  \midrule
  \multirow{4}{*}{\texttt{dexcode}}
    & $S_{5}$ Restricted API calls & \checkmark & \checkmark \\
    & $S_{6}$ Used permission & \xmark & \xmark\\
    & $S_{7}$ Suspicious API calls & \checkmark & \checkmark \\
    & $S_{8}$ Network addresses & \checkmark & \checkmark \\
  \bottomrule
\end{tabular}
\label{tab:map_feature_sets}
\end{table}

\noindent
\textbf{Evasion Attacks Setting.} We randomly select $800$ malware examples from the test set to wage evasion attacks by using the attack algorithms described in Section \ref{sec:threat_model}. In the settings of Random, Grosse, BGA, BCA, and $\ell_1$-PGD attacks, we iterate these algorithms until reaching a predefined maximum number of steps, while noting that Grosse, BGA, and BGA attacks leverage feature addition only. For waging the Mimicry attack, in order to increase its effectiveness, we use $N_b$ benign examples to guide the perturbation of a single malware example, leading to $N_b$ perturbed examples; then, we select a resulting example such that it causes the mis-classification with the smallest perturbation. Therefore, we denote this attack as Mimicry$\times N_b$. For other attacks, we set $\varepsilon = 1.0$ for the FGSM attack. In $\ell_\infty$ norm and Adam based PGD attacks, the step size is $\alpha = 0.01$ with iterative times $100$. The $\ell_2$ norm PGD attack is performed for 100 iterations with step size $1.0$. 

\smallskip

\noindent{\bf Experimental Validation of Functionality}. In order to test whether or not perturbations in the feature space render to executable files in the example space, we use Cuckoodroid \cite{cuckoodroid:ref} to install and run APKs in an Android emulator. Owing to efficiency considerations, we randomly select 10 malware APKs and generate their perturbed APKs using the PGD-Adam attack against the Basic DNN model. Among the 10 original (i.e., unperturbed) APKs, all of them can be loaded but 2 cannot run in the Android emulator. Among the 10 perturbed examples, all of them can be loaded but 5 of them cannot run (and 2 of these 5 correspond to the 2 original APKs that cannot run). This means that more research is needed in order to systematically assure that perturbation can indeed preserve the functionalities of malware examples, which is unique to adversarial malware detection \cite{kucuk2020deceiving,pierazzi2020intriguing}.

\subsection{Experimental Results}

\begin{table}[!htbp]
	\caption{Effectiveness of the defense framework when there are no adversarial attacks.
	}
	\centering
	\begin{tabular}{l|c|c|c}
		\toprule
		{Defense}& {FNR (\%)} & {FPR (\%)} & {Accuracy (\%)}\\
		\midrule
		Basic DNN & 3.684 &	0.320 & {\bf 99.28} \\
		Adversarial Training & 3.246 & 1.777 & 98.05 \\
		Adversarial Regularization & 4.737 & {\bf 0.190} & 99.27 \\
		DAE & 3.246 & 0.450 & 99.22 \\
		AT+DAE & 3.246 & 1.694 & 98.12 \\
		Ensemble AT+DAE & {\bf 2.456} & 2.464 & 97.54 \\
		\bottomrule
	\end{tabular}
	\label{tab:no-attack-res}
\end{table}

\begin{table*}[!htbp]
	\caption{Effectiveness of the defense framework against 
		grey-box adversarial malware evasion attacks.}
	\centering
	\begin{tabular}{l|cccccc}
		\toprule
		\multirow{2}{*}{Attack}&\multicolumn{6}{c}{Accuracy (\%)} \\\cmidrule{2-7}
		& Basic DNN & Adversarial Training (AT) & Adversarial Regularization & DAE & AT+DAE & Ensemble AT+DAE \\\midrule 
		No Attack & 96.63 & 97.00 & 95.63 & 96.88 & 96.50 & \textbf{97.75} \\\hline
		Random Attack & \textbf{100.0} & \textbf{100.0} & \textbf{100.0} & \textbf{100.0} & \textbf{100.0} & \textbf{100.0} \\\hline
		Mimicry$\times 1$ & 53.88 &	86.13 & 52.75 &	56.88 &	91.50 &	\textbf{96.13} \\\hline
		Mimicry$\times 10$ & 35.25 & 85.63 & 34.88 & 52.63 & 85.13 & \textbf{89.88} \\\hline
		FGSM \cite{goodfellow6572explaining} & 4.00 & 97.50 & 95.88 & 96.88 & 96.75 & \textbf{98.00} \\\hline
		Grosse \cite{grosse2017adversarial} & 1.13 & 97.00 & 11.75 & 65.13 & 97.63 & \textbf{99.38} \\\hline
		BGA \cite{al2018adversarial} & 0.25 & \textbf{100.0} & 71.13 & \textbf{100.0} & \textbf{100.0} & \textbf{100.0} \\\hline
		BCA \cite{al2018adversarial} & 0.25 & \textbf{100.0} & 49.50 & 58.00 & \textbf{100.0} & \textbf{100.0} \\\hline
		PGD-$\ell_1$ & 0.25 & \textbf{100.0} & 43.88 & 53.88 & \textbf{100.0} & \textbf{100.0} \\\hline
		PGD-$\ell_2$ & 58.63 & \textbf{100.0} & 99.75 & \textbf{100.0} & \textbf{100.0} & \textbf{100.0} \\\hline
		PGD-$\ell_\infty$ & 0.25 & \textbf{100.0} & \textbf{100.0} & \textbf{100.0} & \textbf{100.0} & \textbf{100.0} \\\hline
		PGD-Adam & 52.50 & \textbf{100.0} &	\textbf{100.0} & \textbf{100.0} & \textbf{100.0} & \textbf{100.0} \\\bottomrule 
	\end{tabular}
	\label{tab:grey-box-attack-res}
\end{table*}

\begin{table*}[!htbp]
	\caption{Effectiveness of the defense framework against 
		white-box adversarial malware evasion attacks.}
	\centering
	\begin{tabular}{l|cccccc}
		\toprule 
		\multirow{2}{*}{Attack}&\multicolumn{6}{c}{Accuracy (\%)} \\\cmidrule{2-7} 
		& Basic DNN & Adversarial Training (AT) & Adversarial Regularization & DAE & AT+DAE & Ensemble AT+DAE \\\midrule 
		Mimicry$\times 10$ & 11.63 & 68.25 & 14.88 & 40.88 & 69.13 & \textbf{79.75} \\\hline
		FGSM \cite{goodfellow6572explaining} & 0.00 & 97.00 & 95.00 & 96.88 & 96.50 & \textbf{97.75} \\\hline
		Grosse \cite{grosse2017adversarial} & 0.00 & 60.75 & 16.63 & 35.50 & 81.13 & \textbf{91.75} \\\hline
		BGA \cite{al2018adversarial} & 0.00 & 97.00 & 91.50 & 74.00 & 96.50 & \textbf{97.50} \\\hline
		BCA \cite{al2018adversarial} & 0.00 &  61.13 & 16.63 & 35.38 &	81.50 &	\textbf{91.75} \\\hline
		PGD-$\ell_1$ & 0.00 & 69.50 & 21.88 & 51.00 & 81.25 & \textbf{88.50} \\\hline
		PGD-$\ell_2$ & 3.00 & \textbf{93.63} & 82.13 & 89.75 & 91.13 & 91.63 \\\hline
		PGD-$\ell_\infty$ & 0.00 &	\textbf{90.38} & 89.75 & 35.38 &	85.50 &	73.63 \\\hline
		PGD-Adam & 1.13  & \textbf{95.13} & 89.63 & 88.25 & 92.88 & 90.00 \\\bottomrule 
	\end{tabular}
	\label{tab:white-box-attack-res}
\end{table*}

\textbf{The Case of No Adversarial Attacks.} Table \ref{tab:no-attack-res} summarizes the classification results on the test set, which are measured with the standard metrics of False Negative Rate (FNR), False Positive Rate (FPR), and classification Accuracy (i.e., the percentage of the test examples that are classified correctly) \cite{Pendleton:2016}.  We observe that when compared with the Basic DNN, Adversarial Training achieves a lower FNR ($0.438\%$ lower) but a higher FPR ($1.457\%$ higher). A similar tendency is exhibited by DAE, AT+DAE and Ensemble AT+DAE. This can be explained as follows: by injecting adversarial malware examples into the training set, the learning process makes the model search for malware examples in a bigger space, explaining the drop in FNR and increase in FPR and therefore a slightly drop ($\leq 1.74\%$) in the classification accuracy. Adversarial Regularization achieves a comparable classification accuracy as Basic DNN,  but the highest FNR among the classifiers we considered. This is caused by the fact that DNN is regularized using small perturbations applied to both benign and malicious samples.
In summary, we draw:
\begin{insight}
In the absence of adversarial attacks, Adversarial Training and DAE can detect more malware examples than the Basic DNN (because of their smaller FNR), at the price of a small side-effect in the FPR and therefore the classification accuracy; Adversarial regularization achieves comparable accuracy as the Baisc DNN while increasing the FNR.
\end{insight}

\noindent
\textbf{The Case of Grey-box Attacks.} Table \ref{tab:grey-box-attack-res} reports the classification results of the defense framework against grey-box attacks. We make the following observations. First, Basic DNN cannot defend against evasion attacks and is completely ruined by attacks that include Mimicry, FGSM, Grosse, BGA, BCA, PGD-$\ell_1$, and PGD-$\ell_\infty$. Second, Adversarial Training significantly enhances the robustness of DNN, achieving the accuracy of 86.13\% and 85.63\% against the Mimicry$\times 1$ and Mimicry$\times 10$ attack respectively and a 100\% accuracy against the other 6 attacks (i.e., BGA, BCA and 4 variants of PGD). Third, Adversarial Regularization, without seeing any adversarial examples, can defend against FGSM, PGD-$\ell_\infty$, PGD-$\ell_2$ and PGD-Adam attacks, but are not effective against attacks such as Grosse, BCA, and PGD-$\ell_1$. A similar phenomenon is observed for DAE. Nevertheless, when using Adversarial Training and DAE together, namely AT+DAE, the defense achieves the highest robustness against evasion attacks than using Adversarial Training and DAE individually, except for the Mimicry$\times 10$ attack and FGSM attack (encountering a \url{~}1\% decrease). Fourth, the Ensemble AT+DAE consists of five AT+DAE classifiers and achieves the highest robustness among the considered defenses against the attacks investigated. In summary, we draw:

\begin{insight}
Under grey-box attack scenario, Adversarial Training is an effective defense against evasion attacks;
DAE offers some defense capability that may not be offered by 
Adversarial Training;
using an ensemble of five AT+DAE classifiers is more effective than using a single AT+DAE classifier against evasion attacks;
Without knowing the attacker's manipulation set, Adversarial Regularization enhances the robustness of Basic DNN but cannot defend attacks such as Grosse.
\end{insight}

\smallskip
\noindent
\textbf{The Case of White-box Attacks.} Table \ref{tab:white-box-attack-res} presents the classification results against white-box attacks. We make the following observations.
(i) All attacks can almost completely evade Basic DNN, but the Mimicry attack is, relatively speaking, less effective because this attack leverages less information about the classifiers than what the other attacks do. 
(ii) Adversarial Training is effective against the FGSM attack, BGA attack and PGD-Adam attack, but not effective against the Grosse attack, BCA attack, and PGD-$\ell_1$ attack because these attacks manipulate a few features when generating adversarial examples and these manipulations are unlikely perceived by Adversarial Training (owing to the fact that Adversarial Training penalizes adversarial spaces searched by Adam optimizer). 
(iii) As expected, Adversarial Regularization is less effective than Adversarial Training. Adversarial Regularization achieves a 91.50\% accuracy against the white-box BGA attack, in contrast to the 71.13\% accuracy against the grey-box BGA attack.
This is counter-intuitive but can be attributed to the fact that Adversarial Regularization may render some gradient-based methods, such as BGA, useless, which is a phenomenon known as
\emph{gradient-masking}  \cite{DBLP:journals/corr/abs-1802-00420,tramer2017ensemble,papernotMGJCS16}.
(iv) AT+DAE achieves considerable robustness against those attacks, with at least an 81.13\% accuracy except for the Mimicry$\times 10$ attack, which defeats the AT+DAE defense because Mimicry can make adversarial malware examples similar to benign ones \cite{7917369}.
(v) The ensemble of AT+DAE defense achieves the highest accuracy against the Mimicry$\times 10$, the Grosse attack and the BCA attack than the other defenses, with about  10\% higher accuracy when compared with the AT+DAE defense. However, the ensemble of AT+DAE achieves lower accuracy than AT+DAE against the PGD-$\ell_2$ attack, the PGD-$\ell_\infty$ attack, and the PGD-Adam attack. This may be caused by the fact that the base model AT+DAE cannot effectively mitigate these attacks.
In summary, we draw:

\begin{insight}
Adversarial Training cannot effectively defend against white-box attacks that were not considered in the training phase; DAE can be useful when adversarial training is not effective; employing ensembles can further improve the robustness against certain white-box attacks. That is, no defenses can defeat all white-box attacks effectively. 
\end{insight}

\section{Application to AICS'2019 Challenge When Knowing  Nothing about Attacks} 
\label{exp}

The challenge is in the context of adversarial malware classification, namely devising evasion-resistant, machine learning based malware classifiers. 
The dataset, including both the training set and the test set, consists of feature vectors extracted from Windows malware examples, each of which belongs to one of the following five classes: {\em Virus}, {\em Worm}, {\em Trojan}, {\em Packed malware}, and {\em AdWare}. 
For each example, the features are collected by the challenge organizer via  dynamic analysis, including the Windows API calls and further processed unigram, bigram, and trigram API calls. The feature names (e.g., API calls) and the class labels are ``obfuscated'' by the challenge organizer as integers, while noting the obfuscation preserves the mapping between the features and the integers representation of them. For example, three API calls are represented by three unique integers, say 101, 102, and 103; then, a trigram API call ``101;102;103'' means a sequence of API calls 101, 102, and 103.  
In total there are 106,428 features.

The test set consists of adversarial examples and non-adversarial examples (i.e., unperturbed malware examples).
Adversarial examples are generated by a variety of perturbation methods, which are not known to the participating teams. However, the ground-truth labels of the test examples are not given to the participating teams. This means that the participating teams cannot calculate the accuracy of their detectors by themselves. Instead, they need to submit their classification results (i.e., labels on the examples in the test set) to the challenge organizer, who will calculate the classification score of each participating team. The Challenge organizer decided to use the Macro F1 score as the classification accuracy metric. 
The Macro F1 score is the unweighted mean of the F1 score \cite{sasaki2007truth} for each class of objects in question (i.e., type of malware in this case). The final score is the Harmonic mean upon the two Macro F1 scores, namely the one for the adversarial examples in the test data and the other for the non-adversarial examples in the test data. Given these two numbers, say $a_1$ and $a_2$, their harmonic mean  $\frac{2a_1a_2}{a_1+a_2}$.

\subsection{Basic Analysis of the AICS'2019 Challenge} \label{sec:basic_ana}

\smallskip
\noindent
\textbf{Is the Training Set Imbalanced?}
The training set consists of 12,536 instances, and the test set consists of 3,133 instances. The training set contains 8,678 instances in class `0', 1,883 instances in class `1', 771 instances in class `2', 692 instances in class `3', and 512 instances in class `4'. We can calculate the maximum ratio between the number of instances in different classes is 16.95, indicating that the training set is highly imbalanced. 
In order to cope with the imbalance in the training set, we use the {\it Oversampling} method to replicate randomly selected feature vectors from a class with a small number of feature vectors. The replication process ends until the number of feature vectors is comparable to that of the largest class (i.e., the class with the largest number of feature vectors), where ``comparable'' is measured by a predefined ratio. In order to see the effect of this ratio, we use a 5-fold cross validation on the training set to investigate the impact of this ratio. The classifier consists of neural networks with two fully-connected layers (each layer having 160 neurons with the ReLU activation function), which are optimized via Adam with epochs 50, mini-batch size 128, learning rate 0.001. The model is selected when achieving the best Macro F1 score on the validation set. Table \ref{tab:imbalanced} shows that the Macro F1 score decreases as the oversampling ratio of minority classes increases. In order to make each mini-batch of training data contain examples from all classes, which would be critical in multiclass classification, our experience suggests us to select the 30\% ratio.

\begin{table}[!htbp]
\caption{Accuracy (\%) and Macro F1 score (\%) are reported with a 95\% confidence interval with respect to the ratio parameter (\%), where `---' means learning a classifier using the original training set.}
		\centering
		\begin{tabular}{ccc}
			\toprule 
			 {Ratio (\%)}&{Accuracy (\%)}&{Macro F1 (\%)}\\
			 \midrule 
             ---&93.20$\pm$1.04&85.52$\pm$1.12 \\
            30 &92.86$\pm$0.75&85.47$\pm$1.04 \\
            40 &92.38$\pm$1.00&84.87$\pm$1.07 \\
            50 &92.21$\pm$0.60&84.87$\pm$1.00 \\
            60 &92.48$\pm$1.12&84.62$\pm$1.01 \\
		    \bottomrule 
		\end{tabular}
		\label{tab:imbalanced}
\end{table}

\begin{figure}[!htbp]
  \begin{subfigure}[b]{0.254\textwidth}
    \includegraphics[width=\textwidth]{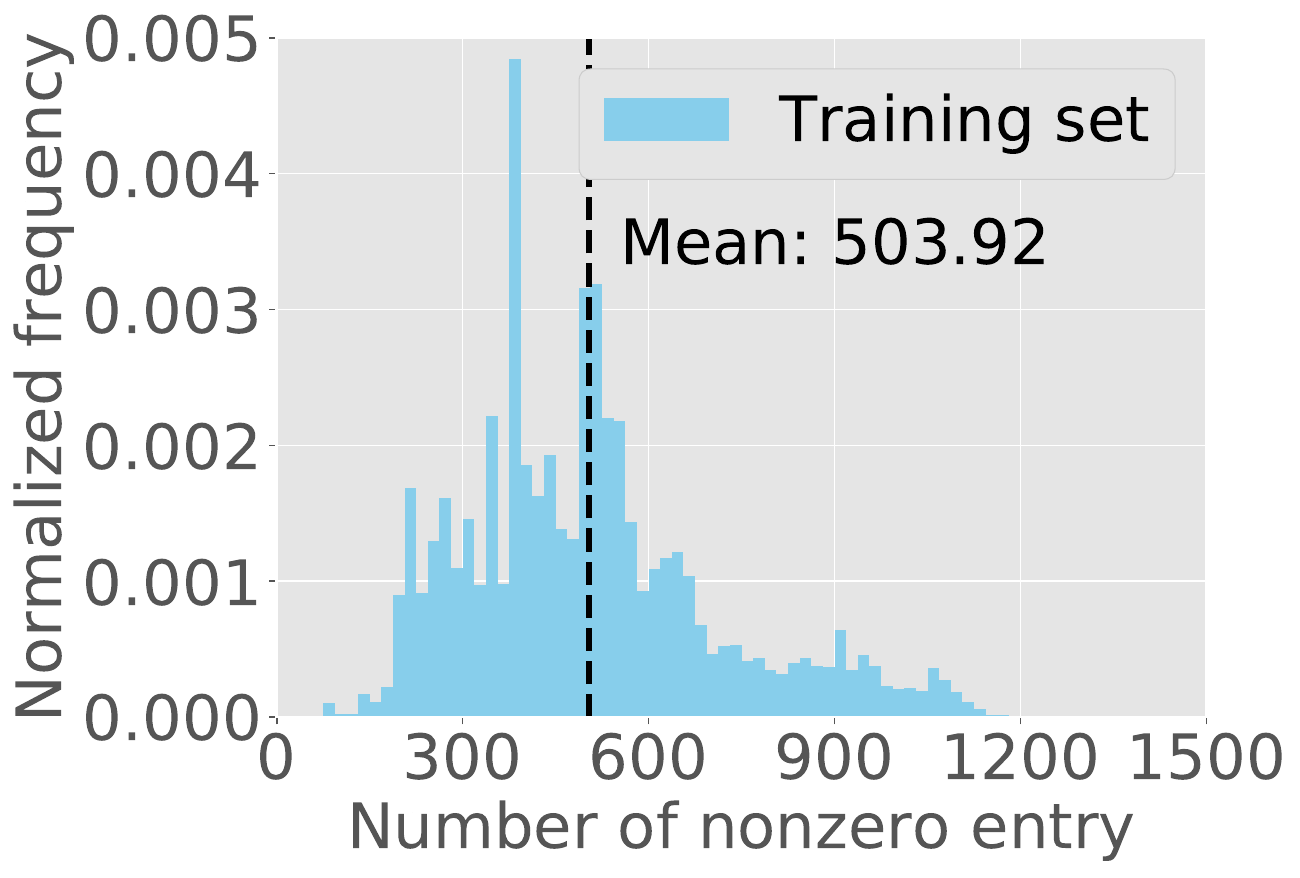}
    \label{fig:train_nonzero}
  \end{subfigure}
  \begin{subfigure}[b]{0.21\textwidth}
    \includegraphics[width=\textwidth]{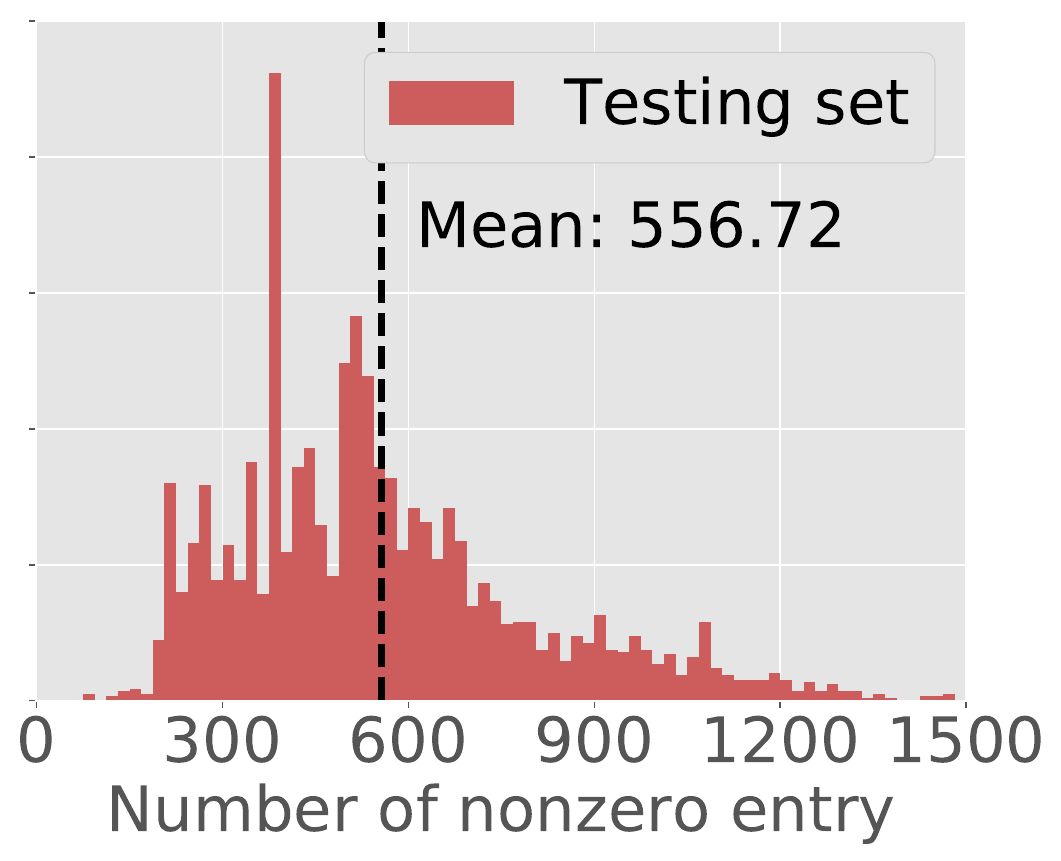}
    \label{fig:test_nonzero}
  \end{subfigure}
\caption{Histogram of the normalized frequency of the number of nonzero entries of feature vectors in the training set vs. test set. The dashed line represents the mean value.}
  \label{fig:nonzero}
\end{figure}

\smallskip
\noindent
\textbf{Are There Simple Indicators of Adversarial Examples?}
In the first test set published by the challenge organizer, we see negative values for some features. These negative values would indicate that they are adversarial examples. In the revised test set provided by the challenge organizer, there are no negative feature values, meaning that there are no simple ways to tell whether an example is adversarial or not. In spite of this, we can speculate on the count of perturbed features by comparing the number of nonzero entries corresponding to feature vectors in the training set and feature vectors in the test set. Figure~\ref{fig:nonzero} shows the normalized frequency of the number of nonzero entries of feature vectors in the training set vs. test set. We observe that their normalized frequencies are similar except that some test examples have more nonzero entries. Their mean values are close and are much smaller than the input dimension ($106,428$), suggesting that the average degree of perturbed features may be small.

\subsection{Classification Results: Challenge Winner}

We train 10 deep neural network classifiers to formulate an ensemble model, including 4 classifiers using the binarization, adversarial regularization, and semantics-preservation techniques discussed in the framework, and the other 6 classifiers using the binarization and adversarial regularization techniques because examples are perturbed without preserving their malicious functionality in the training. Since we do not have access to the malware examples, we cannot tell whether a feature perturbation preserves the malware functionality or not. 
The inner maximization performed by using gradient descent with respect to the transformed input iterates $T=55$ times via the Adam optimizer~\cite{kingmaB14} with learning rate 0.01. We leverage the random start points and $K=5$. The ratio for ensemble of random subspace method is set as $\Lambda=0.5$. Each base classifier has two fully-connected hidden layers (each layer having neurons 160), uses the ELU activation function,
and is optimized by Adam. The ensemble achieves a Macro F1 score of 88.30\% upon non-attack dataset, a 63.0\% Macro F1 score under attacks, and a Harmonic mean on both Macro F1 scores of 73.60\%. This is the highest Harmonic Mean score among the participating teams. Although this score is not ideal, this may be inherent to the fact that we as the defender do not know any information about the attack.
This leads to:
\begin{insight}
The information ``barrier'' that the defender does not know the attacker's manipulation set is a fundamental one because the attacker may use adversarial malware examples that are far away from what the defender would use to train its defense model. 
\end{insight}

\section{Application to AICS'2019 Challenge after Knowing Ground Truth}\label{exp-new}

After the Challenge organizer announced that we won the Challenge, the ground-truth labels of the test set are released so that we can conduct further study. We stress that we still do not know the attacks that were used by the Challenge organizer.

\subsection{Training Classifiers}

\begin{table*}[t]
\caption{Classifiers Accuracy (\%) and Macro F1 score (\%) with no attacks vs. using grey-box adversarial evasion attacks respectively, and the Harmonic mean (\%) of the two Macro F1 scores.}
		\centering
		\begin{tabular}{l|cc|cc|c}
			\toprule 
\multirow{2}{*}{Classifiers}&\multicolumn{2}{c|}{No attacks (\%)}&\multicolumn{2}{c|}{Attacks (\%)} & \multirow{2}{*}{Harmonic mean (\%)} \\\cmidrule{2-5} 
& Accuracy & Macro F1 & Accuracy & Macro F1 & \\\midrule 
Basic DNN & \textbf{96.24}	& \textbf{88.91} & 63.46 & 35.00 & 50.23 \\\hline
Binarization & 95.80 & 87.99 & 63.79 & 35.47 & 50.56 \\\hline
Adversarial Regularization (AR) & 95.66 & 87.98 & 72.02 & 58.93 & 70.58 \\\hline
Binarization+AR & 95.62	& 87.87 & 75.22	& 59.87 & 71.22 \\\hline
Ensemble Binarization+AR & 95.93 & 88.58 & \textbf{76.02} & \textbf{62.95} & \textbf{73.60} \\\bottomrule 
		\end{tabular}
		\label{tab:aics}
\end{table*}

\smallskip
\noindent
\textbf{Classifier.} We consider and compare five classifiers: (\textbf{i}) Basic DNN without incorporating any defense; (\textbf{ii}) hardened DNN incorporating the binarization technique \cite{xu2017feature} (dubbed Binarization); (\textbf{iii}) hardened DNN incorporating adversarial regularization (dubbed Adversarial Regularization); (\textbf{iv}) hardened DNN incorporating Binarization and Adversarial Regularization (dubbed Binarization+AR); (\textbf{v}) an ensemble of Binarization+AR classifiers (dubbed Ensemble Binarization+AR). 

\smallskip
\noindent
\textbf{Hyper-parameter Settings.} All of the DNNs we use have two fully-connected hidden layers (each layer having 160 neurons), use the ReLU activation function,
and are optimized by Adam with epochs 30, mini-batch size 128, and learning rate 0.001. For Adversarial Regularization, we perform the inner maximization via Adam (with learning rate 0.01). Our preliminary experiments suggest us to set iterations $T=60$. The starting point is chosen from $K=5$ initialized points with salt-and-pepper noises, which have a noise ratio $\epsilon^r$ chose uniformly at random from $0$ to $\epsilon_{max}^r=10\%$. This means at most $10\%$ of the features can be changed by salt-and-pepper noises in each training round. For the ensemble, we train 5 Binarization+AR classifiers, each of which is learned from an $80\%$ data randomly selected from the training set, with a $\Lambda=0.5$ fraction of features. We augment the training set for the last three classifiers as described in Section \ref{sec:basic_ana}. 

\subsection{Classification Results}

Table \ref{tab:aics} presents the results with and without adversarial attacks. We make three observations. (i) Adversarial Regularization significantly improves the Macro F1 score against the attacks when compared with the Basic DNN (a $23.93\%$ higher Macro F1 score). The Macro F1 score of Adversarial Regularization in the absence of adversarial attacks drops slightly when compared with the Basic DNN ($\approx 1\%$). (ii) By comparing Binarization (row 2) and the Basic DNN, Binarization can improve the robustness of DNN against adversarial attacks a little bit (a 0.47\% increase in the Macro F1 score). (iii) Ensemble Binarization+AR achieves a higher classification accuracy than Binarization+AR, in the presence or absence of adversarial attacks.

\begin{table}[!htbp]
\caption{Accuracy (\%) and Macro F1 score (\%) of Adversarial Regularization in the absence vs. presence of adversarial evasion attacks, with respect to the maximum salt-and-pepper noise ratio $\epsilon_{max}^r$, where $*$ means that a classifier is learned using oversampling.}
		\centering
		\begin{tabular}{l|cc|cc}
			\toprule
\multirow{2}{*}{Noise Ratio (\%)}&\multicolumn{2}{c|}{No attacks (\%)}&\multicolumn{2}{c}{Attacks (\%)} \\\cmidrule{2-5} 
& Accuracy & Macro F1 & Accuracy & Macro F1 \\\midrule 
$\epsilon_{max}^r=0$ & 96.11 & 88.43 & 69.68 & 49.87 \\\hline
$\epsilon_{max}^r=0.1$ & 95.93 & 88.10 & 74.00 & 50.52 \\\hline
$\epsilon_{max}^r=1$ & {\bf 96.24} & {\bf 89.14} & 73.11 & 55.98 \\\hline
$\epsilon_{max}^r=10$ & 96.19 & 88.46 & {\bf 77.11} & 56.22 \\\hline
$\epsilon_{max}^r=20$ &96.06 & 88.14	& 75.11 & 51.23 \\\hline
$\epsilon_{max}^r=10^\ast$ & 95.66	& 87.98 & 72.02 & {\bf 58.93} \\\bottomrule 
		\end{tabular}
		\label{tab:aics_hp_analysis}
\end{table}

\smallskip
\noindent
\textbf{Hyper-parameters Sensitivity.} In Adversarial Regularization, $\epsilon_{max}^r$ is crucial and is set manually. Intuitively, a greater $\epsilon_{max}^r$ lets the defense perceive a larger space, but inhibiting the convergence of training. In addition, we want to know whether the oversampling is useful or not for Adversarial Regularization. We thus conduct a group of experiments to justify these settings. Table \ref{tab:aics_hp_analysis} shows the experimental results. We observe that the Macro F1 score in the presence of adversarial evasion attacks increase with the increase of $\epsilon_{max}^r$ from $0\%$ to $10\%$. Meanwhile, Accuracy and Macro F1 score do not decrease in the absence of adversarial evasion attacks, and actually slightly increase  at $\epsilon_{max}^r=1\%$. 
Furthermore, when the oversampling technique is leveraged at $\epsilon_{max}^r=10\%$ (the last row), both Accuracy and Macro F1 score in the absence of adversarial evasion attacks decrease slightly ($<1\%$). Nevertheless, the Macro F1 score in the presence of adversarial evasion attacks increases from $56.22\%$ to $58.93\%$. This leads us to draw:
\begin{insight}
Oversampling is not necessary when there are no adversarial evasion attacks, but improves the effectiveness of Adversarial Regularization against adversarial evasion attacks in terms of macro F1 score.
\end{insight}

\subsection{Retrospective Analysis of the AICS'2019 Challenge}

\begin{figure}[!htbp]
	\centering
	\scalebox{0.41}{
	\includegraphics[width=0.9\textwidth]{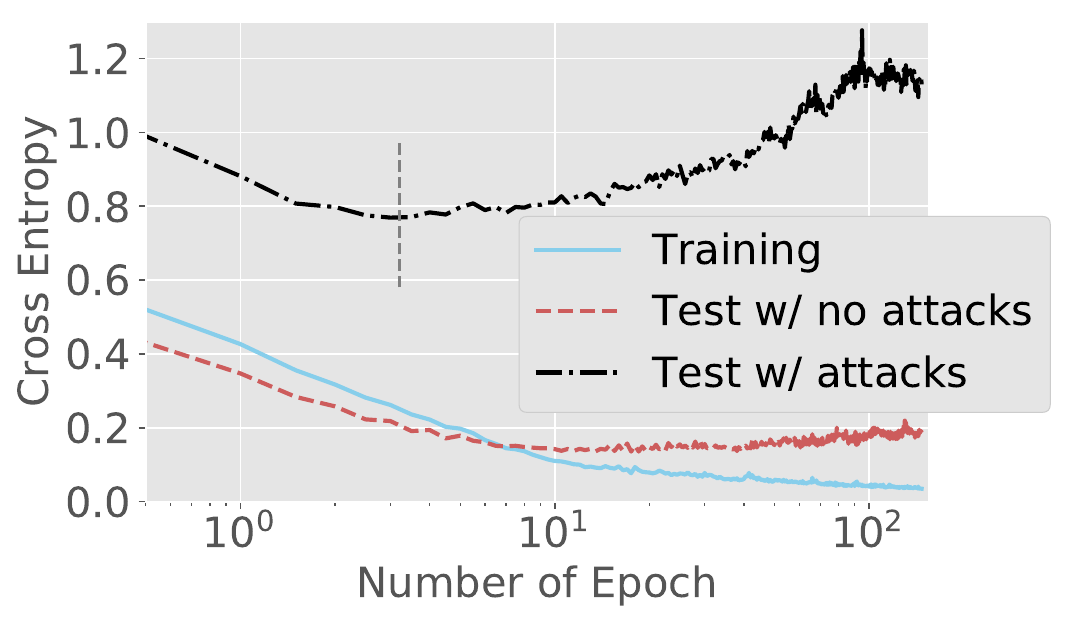}
	}
\caption{Cross entropy loss of the classifier hardened by Adversarial Regularization over the training set, the test set with no adversarial evasion attacks, and the test set with adversarial evasion attacks.}
	\label{fig:over-fit}
\end{figure}

Figure \ref{fig:over-fit} demonstrates that adversarial regularization over-fits the perturbations searched by the inner maximizer unexpectedly. We observe that the cross-entropy loss induced by the perturbations increases significantly after about 10 epochs. Meanwhile, the cross-entropy loss on the test set with no adversarial evasion attacks changes slightly, until the number of epochs approaches 100. This means that the DNN will memorize the perturbations produced in the training phase, leading to poor generalization. Therefore, new defense strategies are needed in order to achieve a much higher accuracy against the Challenge instances. This suggests:
\begin{insight}
Adversarial regularization triggers the over-fitting issue; Without knowing the manipulation set, unsupervised learning may play an important role because unsupervised defenses are devised without using label information about the perturbed examples.
\end{insight}

\section{Conclusion} \label{conclusion}

We have presented six principles for enhancing the robustness of neural network classifiers against adversarial evasion attacks in the setting of malware classification. These principles guided us to design a framework, which is validated via a real-world dataset and the AICS'2019 Challenge. We drew a number of insights that are useful for real-world defenders.

We hope this paper will inspire more research into this important problem. Future research problems are abundant, such as the following.
First, the adversarial training in our study is applied to feature representations satisfying box-constraints (in a discrete space). How should we accommodate other kinds of feature extractions such as graph-based or sequential-like ones \cite{20182405297553,DBLP:conf/kdd/FanHZYA18,cui2020disl,cui2020adversarial}? One possible approach is to instantiate the minmax adversarial training using a generic method, which does not need to know the special knowledge of the hardened model. Second, it is imperative to generate adversarial malware examples in an end-to-end fashion, assuring that a perturbed malware example indeed preserves the functionality of the original, unperturbed malware example. Third, it is an open problem to adapt the provable or certified defense \cite{balunovic2020adversarial} into the context of adversarial malware detection because it is not clear how one should define convex manipulation sets for perturbing malware examples. Unlike the image data where $\ell_p$-norm may quantify visual semantics, $\ell_p$-norm cannot characterize the functionalities of malware examples. 
Fourth, what are the other principles that can be leveraged to defend against adversarial malware examples?

\ifCLASSOPTIONcaptionsoff
  \newpage
\fi

\bibliographystyle{IEEEtran}
\bibliography{malware_paper}

\begin{thebibliography}{10}
\providecommand{\url}[1]{#1}
\csname url@samestyle\endcsname
\providecommand{\newblock}{\relax}
\providecommand{\bibinfo}[2]{#2}
\providecommand{\BIBentrySTDinterwordspacing}{\spaceskip=0pt\relax}
\providecommand{\BIBentryALTinterwordstretchfactor}{4}
\providecommand{\BIBentryALTinterwordspacing}{\spaceskip=\fontdimen2\font plus
\BIBentryALTinterwordstretchfactor\fontdimen3\font minus
  \fontdimen4\font\relax}
\providecommand{\BIBforeignlanguage}[2]{{%
\expandafter\ifx\csname l@#1\endcsname\relax
\typeout{** WARNING: IEEEtran.bst: No hyphenation pattern has been}%
\typeout{** loaded for the language `#1'. Using the pattern for}%
\typeout{** the default language instead.}%
\else
\language=\csname l@#1\endcsname
\fi
#2}}
\providecommand{\BIBdecl}{\relax}
\BIBdecl

\bibitem{DBLP:journals/corr/abs-1812-08108}
\BIBentryALTinterwordspacing
D.~Li, Q.~Li, Y.~Ye, and S.~Xu, ``Enhancing robustness of deep neural networks
  against adversarial malware samples: Principles, framework, and aics'2019
  challenge,'' \emph{CoRR}, vol. abs/1812.08108, 2018. [Online]. Available:
  \url{http://arxiv.org/abs/1812.08108}
\BIBentrySTDinterwordspacing

\bibitem{symantec:Online}
\BIBentryALTinterwordspacing
Symantec. (2018) Symantec {@ONLINE}. [Online]. Available:
  \url{https://www.symantec.com/security-center/threat-report}
\BIBentrySTDinterwordspacing

\bibitem{cisco:Online}
\BIBentryALTinterwordspacing
CISCO. (2018) Cisio {@ONLINE}. [Online]. Available: \url{https://www.cisco.com}
\BIBentrySTDinterwordspacing

\bibitem{DBLP:journals/csur/YeLAI17}
Y.~Ye, T.~Li, D.~A. Adjeroh, and S.~S. Iyengar, ``A survey on malware detection
  using data mining techniques,'' \emph{{ACM} Comput. Surv.}, vol.~50, no.~3,
  pp. 41:1--41:40, 2017.

\bibitem{grosse2017adversarial}
K.~Grosse, N.~Papernot, P.~Manoharan, M.~Backes, and P.~McDaniel, ``Adversarial
  examples for malware detection,'' in \emph{European Symposium on Research in
  Computer Security}.\hskip 1em plus 0.5em minus 0.4em\relax Springer, 2017,
  pp. 62--79.

\bibitem{DBLP:conf/eisic/ChenYB17}
L.~Chen, Y.~Ye, and T.~Bourlai, ``Adversarial machine learning in malware
  detection: Arms race between evasion attack and defense,'' in
  \emph{EISIC'2017}, 2017, pp. 99--106.

\bibitem{al2018adversarial}
A.~Al-Dujaili, A.~Huang, E.~Hemberg, and U.-M. O’Reilly, ``Adversarial deep
  learning for robust detection of binary encoded malware,'' in \emph{2018 IEEE
  Security and Privacy Workshops (SPW)}.\hskip 1em plus 0.5em minus 0.4em\relax
  IEEE, 2018, pp. 76--82.

\bibitem{DBLP:conf/ijcai/HouYSA18}
S.~Hou, Y.~Ye, Y.~Song, and M.~Abdulhayoglu, ``Make evasion harder: An
  intelligent android malware detection system,'' in \emph{Proceedings of the
  Twenty-Seventh IJCAI}, 2018, pp. 5279--5283.

\bibitem{pierazzi2020intriguing}
F.~Pierazzi, F.~Pendlebury, J.~Cortellazzi, and L.~Cavallaro, ``Intriguing
  properties of adversarial ml attacks in the problem space,'' in \emph{2020
  IEEE Symposium on Security and Privacy (SP)}.\hskip 1em plus 0.5em minus
  0.4em\relax IEEE, 2020, pp. 1332--1349.

\bibitem{li2020adversarial}
D.~Li and Q.~Li, ``Adversarial deep ensemble: Evasion attacks and defenses for
  malware detection,'' \emph{IEEE Transactions on Information Forensics and
  Security}, vol.~15, pp. 3886--3900, 2020.

\bibitem{kucuk2020deceiving}
Y.~Kucuk and G.~Yan, ``Deceiving portable executable malware classifiers into
  targeted misclassification with practical adversarial examples,'' in
  \emph{Proceedings of the Tenth ACM Conference on Data and Application
  Security and Privacy}, 2020, pp. 341--352.

\bibitem{szegedyZSBEGF13}
C.~Szegedy, W.~Zaremba, I.~Sutskever, J.~Bruna, D.~Erhan, I.~Goodfellow, and
  R.~Fergus, ``Intriguing properties of neural networks,'' \emph{arXiv preprint
  arXiv:1312.6199}, 2013.

\bibitem{goodfellow6572explaining}
I.~J. Goodfellow, J.~Shlens, and C.~Szegedy, ``Explaining and harnessing
  adversarial examples (2014),'' \emph{arXiv preprint arXiv:1412.6572}.

\bibitem{serban2018adversarial}
A.~C. Serban and E.~Poll, ``Adversarial examples-a complete characterisation of
  the phenomenon,'' \emph{arXiv preprint arXiv:1810.01185}, 2018.

\bibitem{li2020sok}
D.~Li, Q.~Li, Y.~Ye, and S.~Xu, ``Sok: Arms race in adversarial malware
  detection,'' \emph{arXiv preprint arXiv:2005.11671}, 2020.

\bibitem{7917369}
A.~{Demontis}, M.~{Melis}, B.~{Biggio}, D.~{Maiorca}, D.~{Arp}, K.~{Rieck},
  I.~{Corona}, G.~{Giacinto}, and F.~{Roli}, ``Yes, machine learning can be
  more secure! a case study on android malware detection,'' \emph{IEEE
  Transactions on Dependable and Secure Computing}, vol.~16, no.~4, pp.
  711--724, July 2019.

\bibitem{aics_nes:Online}
\BIBentryALTinterwordspacing
M.~Nazir. (2019) Utsa wins global cyber security challenge {@ONLINE}. [Online].
  Available:
  \url{https://www.eurekalert.org/pub_releases/2019-01/uota-uwg011819.php}
\BIBentrySTDinterwordspacing

\bibitem{biggio2010multiple}
B.~Biggio, G.~Fumera, and F.~Roli, ``Multiple classifier systems for robust
  classifier design in adversarial environments,'' \emph{International Journal
  of Machine Learning and Cybernetics}, vol.~1, no. 1-4, pp. 27--41, 2010.

\bibitem{Biggio2010}
------, ``Multiple classifier systems under attack,'' in \emph{International
  Workshop on Multiple Classifier Systems}.\hskip 1em plus 0.5em minus
  0.4em\relax Springer, 2010, pp. 74--83.

\bibitem{smutz2016tree}
C.~Smutz and A.~Stavrou, ``When a tree falls: Using diversity in ensemble
  classifiers to identify evasion in malware detectors.'' in \emph{NDSS}, 2016.

\bibitem{article_stokes}
J.~W. {Stokes}, D.~{Wang}, M.~{Marinescu}, M.~{Marino}, and B.~{Bussone},
  ``Attack and defense of dynamic analysis-based, adversarial neural malware
  detection models,'' in \emph{MILCOM 2018 - 2018 IEEE Military Communications
  Conference (MILCOM)}, 2018, pp. 1--8.

\bibitem{wang_2017}
Q.~Wang, W.~Guo, K.~Zhang, and et~al., ``Adversary resistant deep neural
  networks with an application to malware detection,'' in \emph{Proceedings of
  the 23rd KDD}.\hskip 1em plus 0.5em minus 0.4em\relax ACM, 2017, pp.
  1145--1153.

\bibitem{li2018hashtran}
D.~Li, R.~Baral, T.~Li, H.~Wang, Q.~Li, and S.~Xu, ``Hashtran-dnn: A framework
  for enhancing robustness of deep neural networks against adversarial malware
  samples,'' \emph{arXiv preprint arXiv:1809.06498}, 2018.

\bibitem{YeFOSINT-SI-2018}
L.~Chen, S.~Hou, Y.~Ye, and S.~Xu, ``Droideye: Fortifying security of
  learning-based classifier against adversarial android malware attacks,'' in
  \emph{FOSINT-SI'2018}, 2018, pp. 253--262.

\bibitem{xu2017feature}
W.~Xu, D.~Evans, and Y.~Qi, ``Feature squeezing: Detecting adversarial examples
  in deep neural networks,'' \emph{arXiv preprint:1704.01155}, 2017.

\bibitem{xu2014evasion}
L.~Xu, Z.~Zhan, S.~Xu, and K.~Ye, ``An evasion and counter-evasion study in
  malicious websites detection,'' in \emph{CNS, 2014 IEEE Conference on}.\hskip
  1em plus 0.5em minus 0.4em\relax IEEE, 2014, pp. 265--273.

\bibitem{kurakin2016adversarial}
A.~Kurakin, I.~Goodfellow, and S.~Bengio, ``Adversarial examples in the
  physical world,'' \emph{arXiv preprint arXiv:1607.02533}, 2016.

\bibitem{madry2017towards}
A.~Madry, A.~Makelov, L.~Schmidt, D.~Tsipras, and A.~Vladu, ``Towards deep
  learning models resistant to adversarial attacks,'' \emph{arXiv preprint
  arXiv:1706.06083}, 2017.

\bibitem{drucker1992improving}
H.~Drucker and Y.~Le~Cun, ``Improving generalization performance using double
  backpropagation,'' \emph{IEEE Transactions on Neural Networks}, vol.~3,
  no.~6, pp. 991--997, 1992.

\bibitem{Lyu:2015:UGR:2919336.2920639}
\BIBentryALTinterwordspacing
C.~Lyu, K.~Huang, and H.-N. Liang, ``A unified gradient regularization family
  for adversarial examples,'' in \emph{Proceedings of the 2015 IEEE
  International Conference on Data Mining (ICDM)}, ser. ICDM '15.\hskip 1em
  plus 0.5em minus 0.4em\relax Washington, DC, USA: IEEE Computer Society,
  2015, pp. 301--309. [Online]. Available:
  \url{http://dx.doi.org/10.1109/ICDM.2015.84}
\BIBentrySTDinterwordspacing

\bibitem{miyato2016adversarial}
T.~Miyato, A.~M. Dai, and I.~Goodfellow, ``Adversarial training methods for
  semi-supervised text classification,'' \emph{arXiv preprint
  arXiv:1605.07725}, 2016.

\bibitem{vincent2010stacked}
P.~Vincent, H.~Larochelle, I.~Lajoie, Y.~Bengio, and P.-A. Manzagol, ``Stacked
  denoising autoencoders: Learning useful representations in a deep network
  with a local denoising criterion,'' \emph{Journal of machine learning
  research}, vol.~11, no. Dec, pp. 3371--3408, 2010.

\bibitem{mengc_2017}
D.~Meng and H.~Chen, ``Magnet: a two-pronged defense against adversarial
  examples,'' pp. 135--147, 2017.

\bibitem{aics_challenge:Online}
\BIBentryALTinterwordspacing
M.~Nazir. (2019) Aics 2019 workshop challenge problem. [Online]. Available:
  \url{http://www-personal.umich.edu/~arunesh/AICS2019/challenge.html}
\BIBentrySTDinterwordspacing

\bibitem{Biggio:Evasion}
I.~C. B.~Biggio and D.~M. et~al., ``Evasion attacks against machine learning at
  test time,'' in \emph{Machine Learning and Knowledge Discovery in Databases:
  European Conference}.\hskip 1em plus 0.5em minus 0.4em\relax Springer, 01
  2013, pp. 387--402.

\bibitem{rndic_laskov}
P.~L. Nedim~rndic, ``Practical evasion of a learning-based classifier: A case
  study,'' in \emph{Security and Privacy (SP), 2014 IEEE Symposium on}.\hskip
  1em plus 0.5em minus 0.4em\relax IEEE, 2014, pp. 197--211.

\bibitem{DBLP:journals/corr/RosenbergSRE17}
I.~Rosenberg, A.~Shabtai, L.~Rokach, and Y.~Elovici, ``Generic black-box
  end-to-end attack against rnns and other calls based malware classifiers,''
  \emph{arXiv preprint}, 2017.

\bibitem{Chen:2017:SES}
L.~Chen, S.~Hou, and Y.~Ye, ``Securedroid: Enhancing security of machine
  learning-based detection against adversarial android malware attacks,'' in
  \emph{ACSAC}.\hskip 1em plus 0.5em minus 0.4em\relax USA: ACM, 2017, pp.
  362--372.

\bibitem{dang2017evading}
H.~Dang, Y.~Huang, and E.-C. Chang, ``Evading classifiers by morphing in the
  dark,'' in \emph{CCS}.\hskip 1em plus 0.5em minus 0.4em\relax ACM, 2017, pp.
  119--133.

\bibitem{anderson2017evading}
H.~S. Anderson, A.~Kharkar, B.~Filar, and P.~Roth, ``Evading machine learning
  malware detection,'' \emph{Black Hat}, 2017.

\bibitem{316904628}
W.~Xu, Y.~Qi, and D.~Evans, ``Automatically evading classifiers: A case study
  on pdf malware classifiers,'' in \emph{NDSS}, January 2016.

\bibitem{Hu2017}
W.~Hu and Y.~Tan, ``Generating adversarial malware examples for black-box
  attacks based on gan,'' 02 2017.

\bibitem{carliniW16a}
N.~Carlini and D.~Wagner, ``Towards evaluating the robustness of neural
  networks,'' in \emph{2017 38th IEEE Symposium on Security and Privacy
  (SP)}.\hskip 1em plus 0.5em minus 0.4em\relax IEEE, 2017, pp. 39--57.

\bibitem{papernot_2016}
N.~Papernot, P.~McDaniel, S.~Jha, M.~Fredrikson, Z.~B. Celik, and A.~Swami,
  ``The limitations of deep learning in adversarial settings,'' in
  \emph{Security and Privacy (EuroS\&P), 2016 IEEE European Symposium
  on}.\hskip 1em plus 0.5em minus 0.4em\relax IEEE, 2016, pp. 372--387.

\bibitem{217486}
O.~Suciu, R.~Marginean, Y.~Kaya, H.~D. III, and T.~Dumitras, ``When does
  machine learning {FAIL}? generalized transferability for evasion and
  poisoning attacks,'' in \emph{27th {USENIX} Security Symposium ({USENIX}
  Security 18)}.\hskip 1em plus 0.5em minus 0.4em\relax Baltimore, MD: {USENIX}
  Association, Aug. 2018, pp. 1299--1316.

\bibitem{DBLP:journals/corr/abs-1904-13000}
\BIBentryALTinterwordspacing
F.~Tram{\`{e}}r and D.~Boneh, ``Adversarial training and robustness for
  multiple perturbations,'' \emph{CoRR}, vol. abs/1904.13000, 2019. [Online].
  Available: \url{http://arxiv.org/abs/1904.13000}
\BIBentrySTDinterwordspacing

\bibitem{kingmaB14}
D.~P. Kingma and J.~Ba, ``Adam: {A} method for stochastic optimization,''
  \emph{CoRR}, vol. abs/1412.6980, 2014.

\bibitem{raff2017malware}
E.~Raff, J.~Barker, J.~Sylvester, R.~Brandon, B.~Catanzaro, and C.~Nicholas,
  ``Malware detection by eating a whole exe,'' \emph{arXiv preprint
  arXiv:1710.09435}, 2017.

\bibitem{zhou2012ensemble}
Z.-H. Zhou, \emph{Ensemble methods: foundations and algorithms}.\hskip 1em plus
  0.5em minus 0.4em\relax CRC press, 2012.

\bibitem{709601_rss}
T.~K. Ho, ``The random subspace method for constructing decision forests,''
  \emph{IEEE Transactions on PAMI}, vol.~20, no.~8, pp. 832--844, 1998.

\bibitem{demontis2019adversarial}
A.~Demontis, M.~Melis, M.~Pintor, M.~Jagielski, B.~Biggio, A.~Oprea,
  C.~Nita-Rotaru, and F.~Roli, ``Why do adversarial attacks transfer?
  explaining transferability of evasion and poisoning attacks,'' in \emph{28th
  $\{$USENIX$\}$ Security Symposium ($\{$USENIX$\}$ Security 19)}, 2019, pp.
  321--338.

\bibitem{DBLP:journals/corr/abs-1811-09300}
\BIBentryALTinterwordspacing
E.~Grefenstette, R.~Stanforth, B.~O'Donoghue, J.~Uesato, G.~Swirszcz, and
  P.~Kohli, ``Strength in numbers: Trading-off robustness and computation via
  adversarially-trained ensembles,'' \emph{CoRR}, vol. abs/1811.09300, 2018.
  [Online]. Available: \url{http://arxiv.org/abs/1811.09300}
\BIBentrySTDinterwordspacing

\bibitem{schott2018towards}
L.~Schott, J.~Rauber, M.~Bethge, and W.~Brendel, ``Towards the first
  adversarially robust neural network model on mnist,'' \emph{arXiv preprint
  arXiv:1805.09190}, 2018.

\bibitem{alain2014regularized}
G.~Alain and Y.~Bengio, ``What regularized auto-encoders learn from the
  data-generating distribution,'' \emph{The Journal of Machine Learning
  Research}, vol.~15, no.~1, pp. 3563--3593, 2014.

\bibitem{luong2015effective}
T.~Luong, H.~Pham, and C.~D. Manning, ``Effective approaches to attention-based
  neural machine translation,'' in \emph{Proceedings of the 2015 Conference on
  Empirical Methods in Natural Language Processing}, 2015, pp. 1412--1421.

\bibitem{Daniel:NDSS}
D.~Arp, M.~Spreitzenbarth, M.~Hubner, H.~Gascon, K.~Rieck, and C.~Siemens,
  ``Drebin: Effective and explainable detection of android malware in your
  pocket.'' in \emph{Ndss}, vol.~14, 2014, pp. 23--26.

\bibitem{236234}
A.~Demontis, M.~Melis, M.~Pintor, M.~Jagielski, B.~Biggio, A.~Oprea,
  C.~Nita-Rotaru, and F.~Roli, ``Why do adversarial attacks transfer?
  explaining transferability of evasion and poisoning attacks,'' in \emph{28th
  {USENIX} Security Symposium ({USENIX} Security 19)}.\hskip 1em plus 0.5em
  minus 0.4em\relax Santa Clara, CA: {USENIX} Association, Aug. 2019, pp.
  321--338.

\bibitem{VirusTotal:Online}
\BIBentryALTinterwordspacing
(2018, May) Virustotal. [Online]. Available: \url{https://www.virustotal.com}
\BIBentrySTDinterwordspacing

\bibitem{10.1007/978-3-319-45719-2_11}
M.~Sebasti{\'a}n, R.~Rivera, P.~Kotzias, and J.~Caballero, ``Avclass: A tool
  for massive malware labeling,'' in \emph{Research in Attacks, Intrusions, and
  Defenses}.\hskip 1em plus 0.5em minus 0.4em\relax Cham: Springer
  International Publishing, 2016, pp. 230--253.

\bibitem{Androguard:Online}
\BIBentryALTinterwordspacing
A.~Desnos. (2019) Androguard {@ONLINE}. [Online]. Available:
  \url{https://github.com/androguard/androguard}
\BIBentrySTDinterwordspacing

\bibitem{cuckoodroid:ref}
I.~Revivo and O.~Caspi, ``Cuckoodroid,'' in \emph{Black Hat USA}, Las Vegas,
  NV, Jul. 2017.

\bibitem{Pendleton:2016}
M.~Pendleton, R.~Garcia-Lebron, J.-H. Cho, and S.~Xu, ``A survey on systems
  security metrics,'' \emph{ACM Comput. Surv.}, vol.~49, no.~4, pp. 1--35, Dec.
  2016.

\bibitem{DBLP:journals/corr/abs-1802-00420}
A.~Athalye, N.~Carlini, and D.~A. Wagner, ``Obfuscated gradients give a false
  sense of security: Circumventing defenses to adversarial examples,''
  \emph{CoRR}, vol. abs/1802.00420, 2018.

\bibitem{tramer2017ensemble}
F.~Tram{\`e}r, A.~Kurakin, N.~Papernot, I.~Goodfellow, D.~Boneh, and
  P.~McDaniel, ``Ensemble adversarial training: Attacks and defenses,''
  \emph{arXiv preprint arXiv:1705.07204}, 2017.

\bibitem{papernotMGJCS16}
N.~Papernot, P.~McDaniel, I.~Goodfellow, S.~Jha, Z.~B. Celik, and A.~Swami,
  ``Practical black-box attacks against deep learning systems using adversarial
  examples,'' \emph{arXiv preprint}, 2016.

\bibitem{sasaki2007truth}
Y.~Sasaki \emph{et~al.}, ``The truth of the f-measure,'' \emph{Teach Tutor
  mater}, vol.~1, no.~5, pp. 1--5, 2007.

\bibitem{20182405297553}
\BIBentryALTinterwordspacing
S.~Cui, B.~Xia, T.~Li, M.~Wu, D.~Li, Q.~Li, and H.~Zhang,
  ``\BIBforeignlanguage{English}{Simwalk: Learning network latent
  representations with social relation similarity},'' vol. 2018-January, 2017,
  pp. 1 -- 6. [Online]. Available:
  \url{http://dx.doi.org/10.1109/ISKE.2017.8258804}
\BIBentrySTDinterwordspacing

\bibitem{DBLP:conf/kdd/FanHZYA18}
Y.~Fan, S.~Hou, Y.~Zhang, Y.~Ye, and M.~Abdulhayoglu, ``Gotcha - sly malware!:
  Scorpion {A} metagraph2vec based malware detection system,'' in
  \emph{Proceedings of KDD'2018}, 2018, pp. 253--262.

\bibitem{cui2020disl}
S.~Cui, T.~Li, S.-C. Chen, M.-L. Shyu, Q.~Li, and H.~Zhang, ``Disl: Deep
  isomorphic substructure learning for network representations,''
  \emph{Knowledge-Based Systems}, vol. 189, p. 105086, 2020.

\bibitem{cui2020adversarial}
S.~Cui, Q.~Li, and S.-C. Chen, ``An adversarial learning approach for
  discovering social relations in human-centered information networks,''
  \emph{EURASIP Journal on Wireless Communications and Networking}, vol. 2020,
  no.~1, pp. 1--19, 2020.

\bibitem{balunovic2020adversarial}
M.~Balunovic and M.~Vechev, ``Adversarial training and provable defenses:
  Bridging the gap,'' in \emph{International Conference on Learning
  Representations}, 2020.

\end{thebibliography}

\end{document}